\documentclass[onlinecite,pra,aps,twocolumn,10pt,floatfix,nofootinbib,superscriptaddress]{revtex4-1}

\usepackage[usenames,dvipsnames]{xcolor}
\usepackage{soul}
\usepackage{amsmath}
\usepackage{amsfonts}
\usepackage{txfonts}
\usepackage{amssymb}
\usepackage{xr-hyper}
\usepackage[colorlinks=true,citecolor=cyan,linkcolor=magenta,filecolor=magenta]{hyperref}
\usepackage[capitalize]{cleveref}
\usepackage{graphicx}
\usepackage{bbold}					
\usepackage[makeroom]{cancel}		
\usepackage{multirow}				
\usepackage[normalem]{ulem}         
\usepackage{array}
\usepackage{mathrsfs}
\usepackage{mathtools}
\usepackage[version=4]{mhchem}      
\usepackage{IEEEtrantools}          
\usepackage{dcolumn}                
\usepackage{physics}                
\usepackage{dsfont}                 
\usepackage{enumerate}              
\usepackage[shortlabels]{enumitem}  
\usepackage{verbatim}
\usepackage{orcidlink}
\usepackage[flushleft]{threeparttable}

\usepackage[globalcitecopy]{bibunits}
\defaultbibliography{bib}
\defaultbibliographystyle{apsrev4-1}

\newcommand{\e}{\mathrm{e}}         
\renewcommand{\i}{\mathrm{i}}       

\DeclareMathOperator{\sign}{sign}  	
\DeclareMathOperator{\diag}{diag}  	

\newcommand{\unarypm}{\mathord{\pm}}

\renewcommand{\vec}[1]{{\vb{#1}}}  
\newcommand{\id}{\mathds{1}}        
\newcommand{\mat}[1]{\begin{pmatrix}#1\end{pmatrix}} 

\newcommand{\nunit}[1]{\,\mathrm{#1}} 

\newcommand{\Ham}{\mathcal{H}}      

\newcommand{\LG}{\mathcal{G}}       
\newcommand{\MLG}{\mathcal{M}}      
\renewcommand{\O}{\mathsf{O}}       
\newcommand{\SO}{\mathsf{SO}}       
\newcommand{\Q}{\mathsf{Q}}         
\newcommand{\PN}{\mathsf{P}}        
\newcommand{\Spin}{\mathsf{Spin}}   

\renewcommand{\so}{\mathfrak{so}}   
\newcommand{\spin}{\mathfrak{spin}} 

\newcommand{\kdotp}{$\vec{k}\cdot\vec{p}$} 

\DeclareMathAlphabet{\mathbbold}{U}{bbold}{m}{n}
\def\abs#1{\left|{#1}\right|}      	
\def\imi{\mathrm{i}}				
\def\e#1{\mathrm{e}^{#1}}				

\def\mcT{\mathcal{T}}					
\def\mcP{\mathcal{P}}					
\def\mcK{\mathcal{K}}					

\def\intg{\mathbbold{Z}}					
\def\ztwo{\mathbbold{Z}_2}					

\newcommand{\colorgapzero}{orange}
\newcommand{\colorgapone}{red}
\newcommand{\colorgaptwo}{blue}

\newcommand{\strainvalue}{-0.01}     

\newcolumntype{x}[1]{>{\centering\let\newline\\\arraybackslash\hspace{0pt}}p{#1}}

\newcommand{\addlinespace}{\rule[\normalbaselineskip]{0pt}{0pt}}
\newcommand{\addextralinespace}[1]{\rule[#1\normalbaselineskip]{0pt}{0pt}}
\definecolor{MyOrange}{RGB}{255,180,0}
\definecolor{MyRed}{RGB}{204,0,0}
\definecolor{MyBlue}{RGB}{0,165.75,229.5}
\definecolor{MyIndigo}{RGB}{75,0,130}
\definecolor{MyTan}{RGB}{185,155,126}

\definecolor{MyGreen}{RGB}{0,216.75,0}

\newcommand{\Python}{Python}
\newcommand{\Mathematica}{\emph{Mathematica}}
\newcommand{\WB}{\texttt{WannierBerri}}

\newcommand{\SM}{SM~\cite{SuppMat}}
\newcommand{\SuppData}{Supplementary Data~\cite{SuppData}}

\newcommand{\nocontentsline}[3]{}
\newcommand{\tocless}[2]{\bgroup\let\addcontentsline=\nocontentsline#1{#2}\egroup}

\crefname{section}{Sec.}{Secs.}
\Crefname{section}{Section}{Sections}

\begin{document}
\begin{bibunit}

\title{From triple-point materials to multiband nodal links}

\author{Patrick M. Lenggenhager\,\orcidlink{0000-0001-6746-1387}}\email[corresponding author: ]{lenpatri@ethz.ch}
\affiliation{Condensed Matter Theory Group, Paul Scherrer Institute, 5232 Villigen PSI, Switzerland}
\affiliation{Institute for Theoretical Physics, ETH Zurich, 8093 Zurich, Switzerland}
\affiliation{Department of Physics, University of Zurich, Winterthurerstrasse 190, 8057 Zurich, Switzerland}

\author{Xiaoxiong Liu\,\orcidlink{0000-0002-2187-0035}}
\affiliation{Department of Physics, University of Zurich, Winterthurerstrasse 190, 8057 Zurich, Switzerland}

\author{Stepan S. Tsirkin}
\affiliation{Department of Physics, University of Zurich, Winterthurerstrasse 190, 8057 Zurich, Switzerland}

\author{Titus Neupert\,\orcidlink{0000-0003-0604-041X}}
\affiliation{Department of Physics, University of Zurich, Winterthurerstrasse 190, 8057 Zurich, Switzerland}

\author{Tom\'{a}\v{s} Bzdu\v{s}ek\,\orcidlink{0000-0001-6904-5264}}
\affiliation{Condensed Matter Theory Group, Paul Scherrer Institute, 5232 Villigen PSI, Switzerland}
\affiliation{Department of Physics, University of Zurich, Winterthurerstrasse 190, 8057 Zurich, Switzerland}

\date{\today}

\begin{abstract}
We study a class of topological materials which in their momentum-space band structure exhibit three-fold degeneracies known as triple points. 
Focusing specifically on \texorpdfstring{$\mcP\mcT$}{PT}-symmetric crystalline solids with negligible spin-orbit coupling, we find that such triple points can be stabilized by little groups containing a three-, four- or six-fold rotation axis, and we develop a classification of all possible triple points as type A vs.~type B according to the absence vs.~presence of attached nodal-line arcs.
Furthermore, by employing the recently discovered non-Abelian band topology, we argue that a rotation-symmetry-breaking strain transforms type-A triple points into multiband nodal links.
Although multiband nodal-line compositions were previously theoretically conceived and related to topological monopole charges, a practical condensed-matter platform for their manipulation and inspection has hitherto been missing.
By reviewing the known triple-point materials with weak spin-orbit coupling, and by performing first-principles calculations to predict new ones, we identify suitable candidates for the realization of multiband nodal links in applied strain.
In particular, we report that an ideal compound to study this phenomenon is \texorpdfstring{\ce{Li2NaN}}{Li2NaN}, in which the conversion of triple points to multiband nodal links facilitates largely tunable density of states and optical conductivity with doping and strain, respectively.
\end{abstract}

\maketitle

\emph{Introduction.---} The recent rapid developments in topological band theory of crystals~\cite{Kane:2005,Qi:2008,Kitaev:2009,Ryu:2010,Fu:2011,Wang:2016,Kruthoff:2017,Po:2017,Bradlyn:2017,Schindler:2018,Vergniory:2019,Bouhon:2019} and engineered metamaterials~\cite{Rechtsman:2013,Kane:2014,Susstrunk:2015,Serra-Garcia:2018,Peri:2020} fuelled a fruitful research into topological band degeneracies in semimetals and metals. Currently, \emph{nodal lines} (NLs)~\cite{Burkov:2011,Fang:2016,Yu:2017,Kim:2015} exhibiting quantized Zak-Berry phase~\cite{Zak:1989a,Berry:1984} are arguably the most investigated type of band degeneracy, and were found to form intricate structures including chains, links and knots~\cite{Bzdusek:2016,Yan:2017,Bi:2017,Ezawa:2017}. More recently, three-fold degenerate nodal points~\cite{Zhu:2016,Wang:2017,Ma:2018,Kim:2018,Heikkila:2015,Das:2020}, also called \emph{triple points} (TPs), received attention as peculiar intermediates between Weyl and Dirac points~\cite{Wan:2011,Xu:2015a,Lv:2015,Soluyanov:2015,Young:2012,Yang:2014,Liu:2014a,Son:2013,Huang:2015}. For spin-orbit-coupled (SOC) systems, TPs were classified into type A vs.~type B according to the absence/presence of attached nodal line arcs~\cite{Zhu:2016}, but they were also reported to occur in certain materials with negligible SOC~\cite{Zhang:2017,Zhang:2017b,Xie:2019,Jin:2019,Jin:2020} including Bernal graphite~\cite{Mikitik:2006}.

Curiously, the topological stability of both NLs and TPs is fully captured only if one partitions the energy bands via multiple energy gaps. On the one hand, NLs in \mbox{space-time}-inversion ($\mcP\mcT$) symmetric systems with negligible SOC (i.e., ``spinless'') exhibit Euler and Stiefel-Whitney \emph{monopole charges}~\cite{Fang:2015,Bzdusek:2017,Ahn:2019b,Lee:2020}, which are related to linking with NLs inside adjacent energy gaps~\cite{Ahn:2018,Tiwari:2020}. On the other hand, TPs formed by a crossing of a one-dimensional (1D) and a two-dimensional (2D) irreducible corepresentation (ICR) along a rotation axis are endpoints at which a NL is transferred between two adjacent energy gaps. Multiband nodal structures in $\mcP\mcT$-symmetric spinless systems were recently characterized by \emph{non-Abelian quaternion invariants}~\cite{Wu:2019,Ahn:2019,Bouhon:2020,Yang:2020b,Kang:2020}, which generalize the Zak-Berry-phase quantization. However, practical condensed-matter platforms to manipulate and probe the non-Abelian band topology have remained limited.

Here we investigate the interplay of TPs and non-Abelian band topology, and we suggest a platform to experimentally observe multiband nodal links in TP materials under strain. To obtain these results, we develop a symmetry-based classification of TPs in spinless $\mcP\mcT$-symmetric systems into type A/B [Fig.~\ref{fig:triplepoint_examples}] similar to the SOC case. By employing the novel insights of non-Abelian quaternion charges, we then illuminate a relation between the NL and TP topology; in particular, we show that type-A TPs in spinless $\mcP\mcT$-symmetric models generally evolve into \emph{multiband nodal links} [Fig.~\ref{fig:argument}] under rotation-symmetry-breaking perturbations~\cite{Zhong:2017}. After reviewing the known spinless TP materials and by performing first-principles calculations to identify additional ones, we finally predict that the recently reported triple-point metal \ce{Li2NaN}~\cite{Jin:2019} produces an ideal multiband nodal-link in applied strain, with the node conversion manifested by unusual strain-tunable optical conductivity. Our results suggest a concrete material platform for experimental studies of non-Abelian band topology, and pave way to uncover further links between seemingly unrelated topological band features. 

\emph{Triple points and non-Abelian band topology.---} TPs are symmetry-protected crossings of a 1D and a 2D ICR occurring along rotation-invariant lines in $\vec{k}$-space [Fig.~\ref{fig:argument}(a)]~\cite{Zhu:2016}. It is possible to recast the corresponding 2D ICR as a NL, in which case the TP marks a location where the NL is transferred from one energy gap (i.e., one pair of energy bands) to an adjacent one [Fig.~\ref{fig:argument}(c)]. In this work we consider $\mcP\mcT$-symmetric systems with negligible SOC (i.e., \emph{spinless}), such that $(\mcP\mcT)^2\!=\!+\id$~\cite{Fang:2015}. A mathematical theorem~\cite{Bouhon:2020} then implies a basis in which the Bloch states are real (rather than complex) vectors. Under these circumstances, topological properties of multiband nodal-line compositions are captured within the framework of non-Abelian band topology~\cite{Wu:2019}.

While the Zak-Berry phase $\varphi_\textrm{ZB}$ (quantized by $\mcP\mcT$ to $0$ vs.~$\pi$ on closed paths) of a block of bands remains invariant provided that these bands are separated by an energy gap from \emph{other} groups of bands, the non-Abelian invariant (called \emph{generalized quaternion charge} by Ref.~\onlinecite{Wu:2019}) $\mathfrak{q}\!\in\!\mathsf{Q}_N$ of an $N$-band system is susceptible to closing \emph{any} of the $(N-1)$ energy gaps [labelled $1$ to ($N-1$) according to increasing energy] between consecutive bands. If $\varphi_\textrm{ZB}^j(\gamma) \!=\! \pi$ for some band $j$ on path $\gamma$, then the square $\mathfrak{q}^2(\gamma) \!=\! -1 \!\in\! \mathsf{Q}_N$ is the unique non-trivial element that squares to identity  $+1 \!\in\! \mathsf{Q}_N$~\cite{Tiwari:2020}. For $\mathfrak{q}(\gamma')\!=\!-1$ all energy bands on $\gamma'$ have a trivial Zak-Berry phase, nevertheless there is a topological obstruction corresponding to an overall $2\pi$ rotation of the $N$-frame (the \emph{eigenframe}) spanned by the Bloch states~\cite{Bouhon:2020}.

\emph{Multiband nodal links.---} We argue that if type-A TPs were to arise in spinless $\mcP\mcT$-symmetric system, they must necessarily evolve into multiband nodal links under rotation-symmetry-breaking strain. To understand this phenomenon, consider a loop $\gamma_0$ [green in Fig.~\ref{fig:argument}(c)] that encircles the NL [\colorgaptwo{} in Fig.~\ref{fig:argument}] extending to one side of a TP [yellow in Fig.~\ref{fig:argument}]. To study the stability of the NL, one could compute the Zak-Berry phase $\varphi_\textrm{ZB}(\gamma_0)$ [Fig.~\ref{fig:argument}(c)] assuming a gap between the two bands that meet at the NL. For a type-A TP it must hold that $\varphi_\textrm{ZB}(\gamma_0) \!=\! 0$, because the loop can be slid to the other side of the TP, where it can be trivialized without closing the corresponding energy gap. However, from the perspective of the non-Abelian invariant the path $\gamma_0$ \emph{cannot} be shrunk to a point on the other side of the TP due to the presence of the NL [\colorgapone{} in Fig.~\ref{fig:argument}] in the adjacent energy gap. This suggests a non-trivial value $\mathfrak{q}(\gamma_0)\!\neq\! +1$ characterizing the NL composition around the type-A TP.
\begin{figure}[t!]
    \centering
    \includegraphics{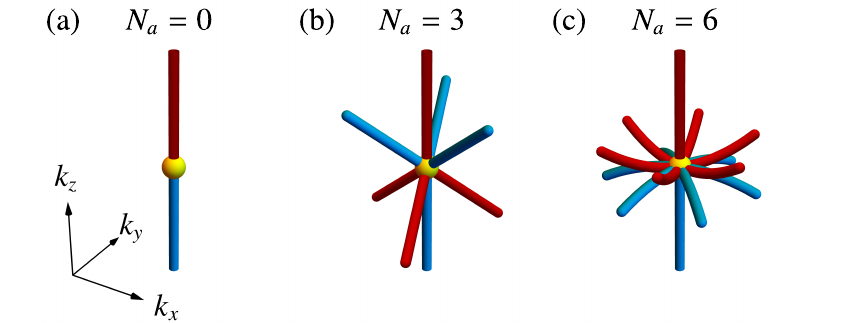}
    \caption{
        Three species of triple points in spinless $\mcP\mcT$-symmetric systems. They are characterized by the number $N_a$ of attached NL arcs, which depends on symmetry as listed in Table~\ref{tab:triplepoint_classif}. We adopt terminology of Ref.~\onlinecite{Zhu:2016}, and call TPs with $N_a\!=\!0$ ($\neq\!0$) type A (type B).
    }
    \label{fig:triplepoint_examples}
\end{figure}

The outlined characterization of the NLs is consistent with the 2D ICR being a \emph{double} nodal line, i.e., one with a quadratic splitting of the two energy bands away from the high-symmetry line~\cite{Xie:2019}. In that case, the Zak-Berry phase on $\gamma_0$ would be trivial, but the quaternion charge would exhibit the non-trivial value $\mathfrak{q}(\gamma_0) \!=\! -1$~\cite{Wu:2019}. Crucially, the non-Abelian invariant remains well-defined even after the 2D ICR is split by a rotation-symmetry-breaking perturbation [Fig.~\ref{fig:argument}(b)], when the TP is lost. While the $\ztwo$-valued Zak-Berry phase $\varphi_\textrm{ZB}(\gamma_0) \!=\! 0$ cannot be directly applied to predict the fate of the NL composition upon such a symmetry breaking, the value $\mathfrak{q}(\gamma_0) \!=\! -1$ implies that $\gamma_0$ \emph{must} enclose a pair of NLs in some energy gap. It therefore follows that the symmetry breaking transforms the TP into a multiband nodal link extending around the Brillouin zone [Fig.~\ref{fig:argument}(d)]. The topological stability of the pair of NLs implies that they carry parallel orientation, which is consistent with the noncommutativity of the non-Abelian invariant~\cite{Tiwari:2020}.
\begin{figure}[t!]
    \centering
    \includegraphics{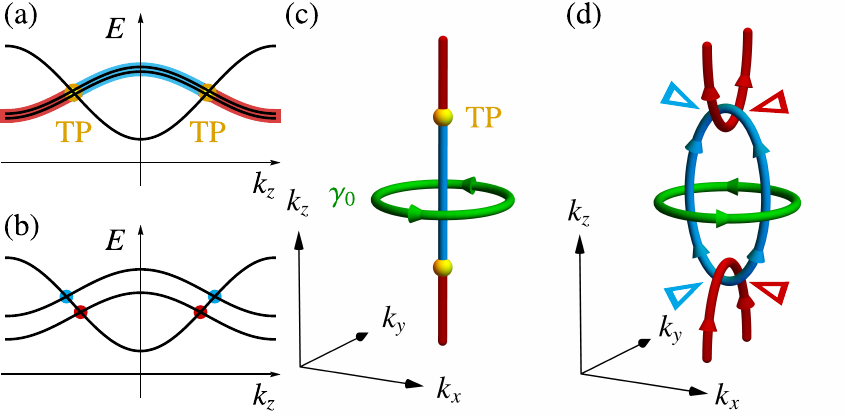}
    \caption{
        Relation of type-A triple points (TPs) to multiband nodal links. Nodal lines (NLs) in the first (second) band gap are displayed in \colorgapone{} (\colorgaptwo{}) and TPs in yellow. ({\bf{a,b}}) Band structure along a high-symmetry line, and ({\bf{c,d}}) the NL composition near the TPs before resp.~after applying rotation-symmetry-breaking perturbation. The green path $\gamma_0$ carries quaternion charge $-1$. The orientations of the NLs (with reversals indicated by triangles of the corresponding color) follow the convention of Ref.~\onlinecite{Tiwari:2020}.
    }
    \label{fig:argument}
\end{figure}
\begin{table}[b!]
\begin{threeparttable}
    \setlength{\belowcaptionskip}{0pt}
    \caption{
        Result of the classification of triple points (TPs) in spinless $\mcP\mcT$-symmetric models (assumed to be non-magnetic and symmorphic). The TP species [characterized by the winding number $w_{\mathrm{2D}}$ of the 2D irreducible corepresentation (ICR), and the number $N_a$ of attached NL arcs per gap] depends on the little group $\LG$. The TP is type A if $N_a\!=\!0$ (and type B otherwise). For $C_{6(v)}$ the result further depends on the pair of intersecting ICRs, where $i\!\in\!\{1,2\}$ for $C_{6v}$. The notation of the ICRs follows Ref.~\onlinecite{Bradley:1972}. TPs with $\abs{w_{\mathrm{2D}}}\!=\!2$ carry quaternion charge $\mathfrak{q}\!=\!-1$, and when $N_a\!=\!0$, they transform to multiband nodal links under strain. The last column reviews previously reported light-element TP materials.
    }
    \label{tab:triplepoint_classif}
    \begin{ruledtabular}
	\begin{tabular}{lcddcl}
        $\LG$       & Pairs of ICRs    & \multicolumn{1}{c}{$\!\abs{w_{\mathrm{2D}}}\!$}	& \multicolumn{1}{c}{$N_a$} &  type & Material candidates    \\
        \hline\addlinespace															
        $C_3$       & (any 2D+1D)         & 1                                      		& 3  & B                       &    \ce{MgH2O2}$^\textrm{a}$                       											\\
        $C_{3v}$    & (any 2D+1D)         & 1                                     		& 3          & B               & Bernal graphite~\cite{Mikitik:2006}        	
        \\
        \addlinespace																																								
        $C_4$       & (any 2D+1D)         & 2                                 			& 0     & A                    &       -                    											\\
        $C_{4v}$    & (any 2D+1D)         & 2                                 			& 0     & A                     &  \ce{ZrO}~\cite{Zhang:2017b}, \ce{Sc3GaC}~\cite{Xie:2019}, \ce{Na2LiN}$^\textrm{a}$                           											\\
        \addlinespace																																								
        $C_6$       & $(E_2,A),(E_1,B)$        & 2                                    			& 0     & A                    &      -                     											\\
        $C_6$       & $(E_1,A),(E_2,B)$      & 2                                    			& 6      & B                   &    \ce{C3N4}$^\textrm{a}$                      											\\
        $C_{6v}$    & $\!(E_1,A_i),(E_2,B_i)\!$         & 2                                    			& 0  & A                       & \ce{Li2NaN}~\cite{Jin:2019} , \ce{TiB2}~\cite{Zhang:2017}             \\
        $C_{6v}$    & $\!(E_2,A_i),(E_1,B_i)\!$         & 2                                    			& 6   & B                      &     \ce{Na3N}~\cite{Jin:2020}                
    \end{tabular}
    \end{ruledtabular}
    \begin{tablenotes}[flushleft]
        \item[a] Compound reported and inspected in Ref.~\onlinecite{Lenggenhager:2022:TPClassif}.
    \end{tablenotes}
    \end{threeparttable}
\end{table}
\begin{figure*}
    \centering
    \includegraphics{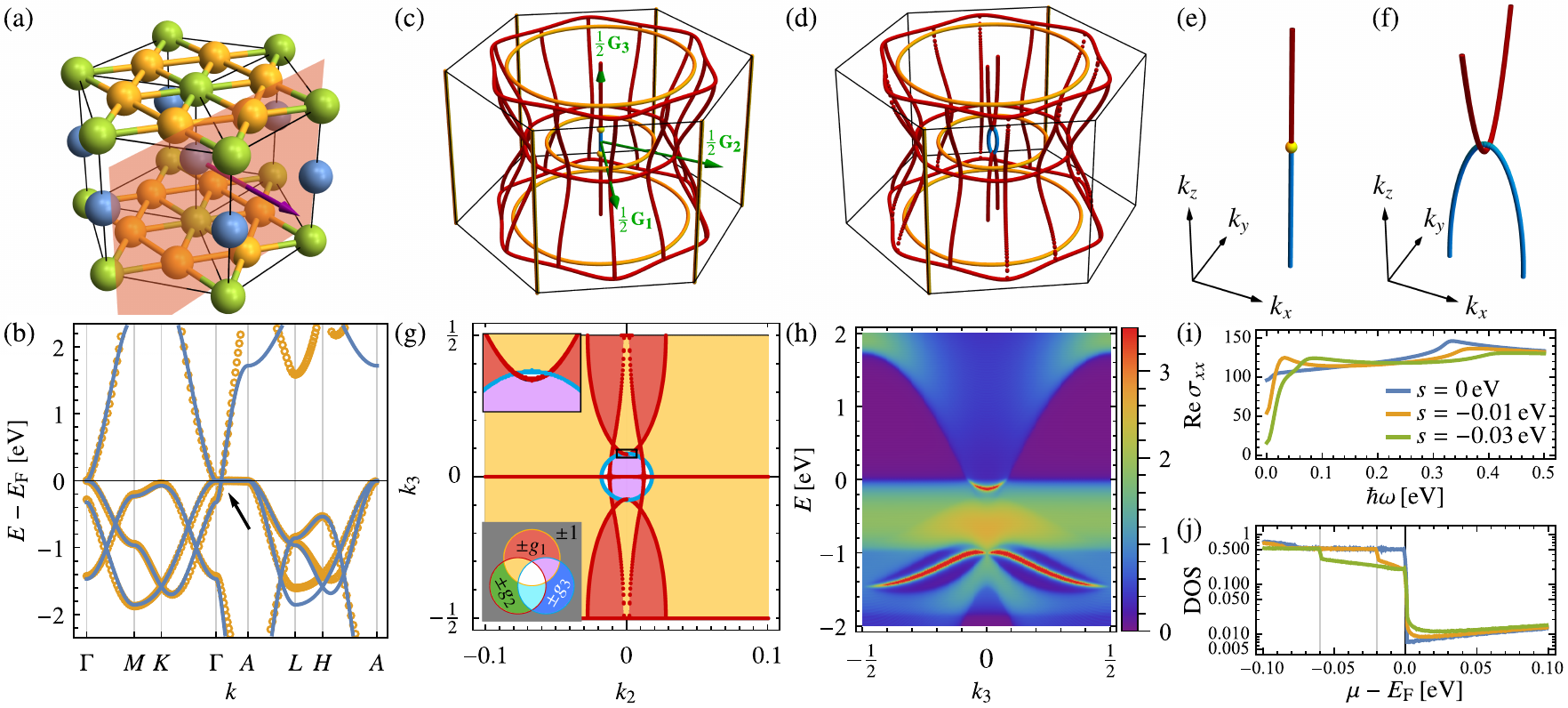}
    \caption{
        \ce{Li2NaN}.
        ({\bf{a}}) Crystal structure with \ce{Na}/\ce{Li}/\ce{N}  (blue/yellow/green) atoms. The red plane indicates the surface termination of (g,h) and the purple arrow its normal. ({\bf{b}}) Band structure along high-symmetry lines: first-principles data (orange points) and fitted four-band tight-binding model (blue lines). The black arrow marks the triple point. ({\bf{c,d}}) Nodal lines (NLs) inside the Brillouin zone without resp.~with strain, and ({\bf{e,f}}) close-ups for $0\!<\!k_z\!<\!0.15\,G_{3}$, colored \colorgapzero{}/\colorgapone{}/\colorgaptwo{} for the first/second/third energy gap. Green arrows in (c) $\vec{G}_{1,2,3}$ indicate primitive reciprocal vectors. ({\bf{g}}) Projection of the NLs in (d) onto the surface Brillouin zone, and the quaternion charge in each region (see main text). Coordinates $k_{2,3}$ and their units correspond to projections of $\vec{G}_{2,3}$. ({\bf{h}}) Surface spectral function for $k_2\!=\!0$. ({\bf{i,j}}) Interband optical conductivity $\textrm{Re}\,\sigma_{xx}$ in $\mathrm{S/cm}$ and density of states (DOS) in $\mathrm{eV^{-1}}$ computed from the fitted tight-binding model for the indicated values of strain $s$~\cite{SuppMat}. In (d,f--h) we set $s\!=\!\strainvalue{}\nunit{eV}$, which corresponds to tensile strain in $y$-direction of $\approx\!1\%$.
    }
    \label{fig:li2nan}
\end{figure*}

\emph{Classification of TPs in spinless band structures.---} The discussed relation between TPs and multiband nodal links relies on (\emph{i}) the 2D ICR carrying a non-trivial quaternion charge, $\mathfrak{q}\!=\!-1$, and (\emph{ii}) the TP being type A. We therefore develop a classification of symmetry-protected TPs in spinless $\mcP\mcT$-symmetric systems according to those two aspects. We restrict our analysis to non-magnetic space groups (SGs), and to high-symmetry lines whose ICRs are not affected by a possible non-symmorphicity of the SG.

To obtain the classification, we consider a rotation-invariant line in $\vec{k}$-space (without loss of generality along $k_z$) of some SG and denote its little group by $\LG$; then, under the above restrictions, $\LG$ is a non-magnetic point group with a single rotation axis and without inversion. By assumption the system is $\mcP\mcT$-symmetric, such that the relevant symmetry group is enhanced to the non-unitary $\MLG \!=\! \LG \cup (\mcP\mcT)\LG$. The stability of a TP requires $\MLG$ to support both 1D and 2D ICRs, which is the case for $\LG\!\in\!\{C_{3(v)},C_{4(v)},C_{6(v)}\}$~\cite{Bradley:1972}. To classify the possible TPs we construct minimal two-band (three-band) \kdotp{} models for all 2D ICRs (combinations of 1D+2D ICRs) of each $\LG$ mentioned above. For that we obtain the necessary IRs from the Bilbao crystallographic server~\cite{Aroyo:2006a,Aroyo:2006b,Elcoro:2017}, then construct the corresponding ICRs~\cite{Bradley:1972} and finally obtain the models using the Python package \texttt{kdotp-symmetry}~\cite{Gresch:PhD}. The constructed models are available in the \SuppData{}.

For two-band spinless $\mcP\mcT$-symmetric models the Hamiltonian is of the form $\Ham(\vec{k})\!=\!h_x(\vec{k})\sigma_x+h_z(\vec{k})\sigma_z$~\cite{Fang:2015}, therefore we can define the winding number $w_{2\textrm{D}}\!\in\!\intg$ of the vector $\vec{h}\!=\!(h_x,h_z)$~\cite{Burkov:2011} for a closed loop $\gamma_0$ [Fig.~\ref{fig:argument}(c)] around the NL corresponding to the 2D ICR. While the winding number is unstable upon the addition of trivial bands, partial topological information is preserved in the quaternion charge. In particular, there is a reduction $w_{2\textrm{D}} \!=\! 2\mod 4 \Rightarrow \mathfrak{q}(\gamma_0)\!=\!-1$ corresponding to the stability of a $2\pi$-rotation of the eigenframe~\cite{Bouhon:2020}. The winding number can be analytically calculated for all the constructed two-band \kdotp{} models. 

Finally, using methods detailed in the companion paper~\cite{Lenggenhager:2022:TPClassif}, we characterize the TPs according to the number $N_a$ of additional attached NLs (which we call \emph{NL arcs}) per band gap. Following the terminology of Ref.~\onlinecite{Zhu:2016} for the SOC case, we call TPs with $N_a\!=\!0$ ($\neq\!0$) type A (type B). In particular, we analyze the NL compositions of the three-band models, which are fully determined by real roots of the discriminant of the characteristic polynomial of the Hamiltonian~\cite{Yang:2020}. For a given group $\LG$ the possible combinations of ICRs fall into at most two topologically distinct equivalence classes.

\emph{Multiband nodal link in strained \ce{Li2NaN}.---} Hexagonal \ce{Li2NaN} (ICSD \#92308) [cf.~Table~\ref{tab:triplepoint_classif}] was reported to be a `superior' TP material with negligible SOC~\cite{Jin:2019}. In particular, it exhibits a pair of type-A TPs at the Fermi level which we find to carry $\mathfrak{q}\!=\!-1$, and no other coexisting Fermi surfaces. The material has a centrosymmetric hexagonal crystal structure with SG $P6/mmm$ (No.~191) [Fig.~\ref{fig:li2nan}(a)]. Additionally, the corresponding 2D ICR has a nearly flat dispersion along the rotation axis [Fig.~\ref{fig:li2nan}(b) path $\Gamma A$], which results in nearly hundredfold change in the density of states within a narrow energy range around the Fermi level~[Fig.~\ref{fig:li2nan}(j)]~(cf.~Supplemental Material (SM)~\cite{SuppMat}). We therefore propose \ce{Li2NaN} as an ideal candidate to study the conversion of the TP into a multiband nodal link, and we argue this conversion to be manifested in sensitive strain-tunable optical conductivity as discussed below.

We employ first-principles calculations to study \ce{Li2NaN}. The bands close to the Fermi level come mainly from nitrogen $p_{x,y,z}$ and sodium $s$ orbitals; thus, we construct a corresponding four-band tight-binding model that reproduces the low-energy theory of \ce{Li2NaN} [Fig.~\ref{fig:li2nan}(b)] after being fitted to first-principles data~(see \SM{}). We find that the vertical NL at $k_x\!=\! 0 \!=\! k_y$ (i.e., the 2D ICR) is well isolated from other NLs in the Brillouin zone (BZ), both in energy and in momentum [Fig.~\ref{fig:li2nan}(b,c)]. The little group along $\Gamma A$ is $C_{6v}$ and the involved ICRs are $(E_1,A_1)$, such that Table~\ref{tab:triplepoint_classif} predicts $N_a \!=\! 0$ and the formation of a multiband nodal link after breaking the $C_{6v}$-symmetry down to $C_{2v}$. In the tight-binding description we model this by adding an energy offset $\delta E \!=\! \pm s$ to nitrogen $p_{x,y}$ orbitals. In agreement with Fig.~\ref{fig:argument}(c,d), we find that the central NL splits into linked NLs in adjacent band gaps [Fig.~\ref{fig:li2nan}(c-f)] with width in $k_{x/y}$-direction proportional to $\sqrt{s}$.

We next examine the bulk-boundary correspondence, which has not yet been clarified for the quaternion charges. We consider a finite slab with the surface termination indicated in Fig.~\ref{fig:li2nan}(a). This corresponds to a projection of the BZ along the reciprocal vector $\vec{G}_1$ [Fig.~\ref{fig:li2nan}(c)] onto the surface Brillouin zone (SBZ). Each point in the SBZ corresponds to a closed path in the BZ, carrying quaternion charge $\mathfrak{q}(k_2,k_3)$ [Fig.~\ref{fig:li2nan}(g)]. Recall~\cite{Tiwari:2020} that any element of $\Q_N$ is given (up to sign) by an ordered product of some of the $(N-1)$ generators $g_j$, where $g_j$ represents band inversion of energy gap $j$ [cf.~legend to Fig.~\ref{fig:li2nan}(g)]. We observe in the surface spectrum along $k_2\!=\!0$ two clear surface bands [Fig.~\ref{fig:li2nan}(h)]. Additional surface bands that hybridize with the bulk states in the second gap are not discernible for small strain. A careful analysis (see \SM{}) reveals that the surface states appear exactly in the gaps with a band inversion according to the value of $\mathfrak{q}$, when including those that have hybridized with bulk states.

The NLs at the Fermi level and their splitting into multiband nodal links suggests non-trivial transport signatures in \ce{Li2NaN}. Using the Kubo-Greenwood formula~\cite{Kubo:1957,Greenwood:1958} which we have implemented in the Python package \WB{}~\cite{Tsirkin:2020,Liu:2020}, we compute the interband (i.e., optical) conductivity $\sigma_{\alpha\beta}(\omega)$ [Fig.~\ref{fig:li2nan}(i)] with and without strain for the fitted tight-binding model. We observe that application of strain leads to a strong suppression of $\Re\sigma_{xx}$ for infrared frequencies $\hbar\omega\!\lesssim\!2\!\abs{s}$. This results from a suppression of optical transitions due to strain-induced mismatch between the chemical potential and the energy of the NLs~\cite{Barati:2017}, because at fixed filling the applied strain moves the blue (red) NL in Fig.~\ref{fig:li2nan}(e,f) above (below) the chemical potential. The strain-dependence of other components of the optical conductivity tensor is presented in \SM. 

Besides the interband contributions to conductivity, in practice one may need to also consider the intraband (Boltzmann) conductivity. While the material's relaxation time which determines this contribution is inherently difficult to predict from first principles, the intraband conductivity is dependent on the density of states at the Fermi level. Since the latter quantity does not exhibit a noticeable growth in applied strain [cf.~Fig.~\ref{fig:li2nan}(j)], we expect the strain-induced suppression of low-frequency conductivity to persist in such a more complete analysis.

\emph{Conclusions and outlook.---} We enrich the library of catalogued band-structure nodes by classifying TPs in spinless systems as type A/B, and by predicting via non-Abelian band topology the conversion of type-A TPs into multiband nodal links in symmetry-breaking strain. We argue that for `ideal' TP materials (i.e.,~with non-dispersive 2D ICR at the Fermi level along the rotation axis, such as predicted in \ce{Li2NaN}), this conversion facilitates a rapidly strain-tunable optical conductivity. We anticipate the non-Abelian band topology to be analogously applicable to study conversions of other species of band nodes, including nodal chains and type-B TPs. This strategy has very recently been exemplified by the predicted Weyl-points-to-nodal-ring conversion in \ce{ZrTe}~\cite{Bouhon:2020,Sun:2018}. We emphasize the generality of this framework, as the ``spinlessness'' condition $(\mcP\mcT)^2\!=\!+\id$ extends to photonic, phononic and magnonic bands.

We expect the non-trivial linking to furnish the here-discussed NL-rings with monopole charges~\cite{Fang:2015,Bzdusek:2017,Ahn:2019b,Lee:2020,Ahn:2018,Tiwari:2020}. In particular, braiding rules of Ref.~\onlinecite{Ahn:2019b} suggest that the `red' NL of \ce{Li2NaN} in Fig.~\ref{fig:li2nan}(e,f) should carry Euler class $\chi\!=\!2$. Note that nodal rings with monopole charges were theoretically related to higher-order topology with hinge Fermi arcs~\cite{Lin:2018,Wang:2019,Wieder:2020}, yet suitable material candidates at present remain very scarce~\cite{Lee:2020}. Our findings suggest that hinge Fermi arcs may arise in spinless triple-point materials; therefore, identifying general conditions for the realization of non-trivial Euler or Stiefel-Whitney monopole charges in TP materials suggests an interesting future application of our results.

\emph{Acknowledgments.---} 
We thank A.~A.~Soluyanov for valuable discussions and to Q.S.~Wu for providing early data that facilitated our work on this project. We further acknowledge discussions and helpful comments on the manuscript from F.~O.~von Rohr, R.-J.~Slager, L.~Trifunovic, and N.~A.~Winter. P.~M.~L. and T.~B. were supported by the Ambizione grant No.~185806 by the Swiss National Science Foundation. T.~N. and S.~S.~T. acknowledge support from the European Research Council (ERC) under the European Union’s Horizon 2020 research and innovation programm (ERC-StG-Neupert-757867-PARATOP) and from the NCCR MARVEL funded by the Swiss National Science Foundation. S.~S.~T. is additionally supported by the grant No.~PP00P2\_176877 by the Swiss National Science Foundation. X.~X.~L was supported by the China Scholarship Council (CSC).

\end{bibunit}
\let\oldaddcontentsline\addcontentsline     
\renewcommand{\addcontentsline}[3]{}        
\bibliography{bib}
\bibliographystyle{apsrev4-1}
\let\addcontentsline\oldaddcontentsline     


\clearpage

\begin{bibunit}
\onecolumngrid
\renewcommand\thesection{\Roman{section}}
\renewcommand\thesubsection{\Alph{subsection}}
\setcounter{page}{1}

\setcounter{figure}{0}
\setcounter{equation}{0}
\setcounter{table}{0}

\renewcommand{\theequation}{S\arabic{equation}}
\renewcommand{\thefigure}{S\arabic{figure}}
\renewcommand{\thetable}{S\arabic{table}}
\renewcommand*{\theHsection}{S.\the\value{section}}

\title{Supplemental Material to:\texorpdfstring{\smallskip \\}{} From triple-point materials to multiband nodal links}

\author{Patrick M. Lenggenhager\,\orcidlink{0000-0001-6746-1387}}\email[corresponding author: ]{lenpatri@ethz.ch}
\affiliation{Condensed Matter Theory Group, Paul Scherrer Institute, 5232 Villigen PSI, Switzerland}
\affiliation{Institute for Theoretical Physics, ETH Zurich, 8093 Zurich, Switzerland}
\affiliation{Department of Physics, University of Zurich, Winterthurerstrasse 190, 8057 Zurich, Switzerland}

\author{Xiaoxiong Liu\,\orcidlink{0000-0002-2187-0035}}
\affiliation{Department of Physics, University of Zurich, Winterthurerstrasse 190, 8057 Zurich, Switzerland}

\author{Stepan S. Tsirkin\,\orcidlink{0000-0001-5064-0485}}
\affiliation{Department of Physics, University of Zurich, Winterthurerstrasse 190, 8057 Zurich, Switzerland}

\author{Titus Neupert\,\orcidlink{0000-0003-0604-041X}}
\affiliation{Department of Physics, University of Zurich, Winterthurerstrasse 190, 8057 Zurich, Switzerland}

\author{Tom\'{a}\v{s} Bzdu\v{s}ek\,\orcidlink{0000-0001-6904-5264}}
\affiliation{Condensed Matter Theory Group, Paul Scherrer Institute, 5232 Villigen PSI, Switzerland}
\affiliation{Department of Physics, University of Zurich, Winterthurerstrasse 190, 8057 Zurich, Switzerland}

\date{\today}

\maketitle
\onecolumngrid

\tableofcontents

\tocless{\section*{Overview}}

In this Supplemental Material we provide supporting information related to our numerical methods, and concerning the triple points (TPs) and the multiband nodal-line (NL) links of \ce{Li2NaN} without and with strain, respectively. 
The information is organized into four parts.
We begin in \cref{SM:Sec:tb_model} with our construction of a four-band tight-binding (TB) model of the compound fitted to first-principles calculations. 
Here, we first summarize in \cref{SM:Subsec:tb_model:dft} the methods employed in the first-principles modelling of \ce{Li2NaN}, and then in \cref{SM:Subsec:tb_model:tb} we present a symmetry-based detachment scheme that allows us to separate the four most relevant bands from the rest of the spectrum.
The fitted TB model has been made available as part of the Supplementary Data~\cite{SuppData}. 

We continue in \cref{SM:Sec:NL_finding} with presenting the algorithm used to find the nodal lines in the full Brillouin zone (BZ) [cf.\ \cref{fig:li2nan}(c--f) of the main text] for the fitted TB model.
Next, in \cref{SM:Sec:tb_model} we carefully analyze the surface states of \ce{Li2NaN} without and with strain [cf.\ \cref{fig:li2nan}(h) of the main text].
We compare them against the bulk quaternion charges [cf.\ \cref{fig:li2nan}(g) of the main text] and observe a consistency with a bulk-boundary correspondence conjectured by Ref.~\onlinecite{Wu:2019} after also including the surface states that hybridize with the bulk states. 
In \cref{SM:Sec:DOS_cond} we present additional data showing the imprint of the strain-induced nodal conversion in \ce{Li2NaN} on its bulk density of states and optical conductivity.
Finally, in \cref{SM:Sec:TP_materials}, we present the first-principles data showing the TPs in the newly proposed material candidates in \cref{tab:triplepoint_classif} and justifying their placement in that table.\bigskip\bigskip

\section{Construction of a tight-binding model for \texorpdfstring{\NoCaseChange{\ce{Li2NaN}}}{Li2NaN}}\label{SM:Sec:tb_model}

In the main text we discuss the triple-point metal \ce{Li2NaN} as a model material which we predict to exhibit the conversion of TPs to multiband nodal links, and we study certain associated material properties.
The results discussed there are obtained from a TB model, whose construction we describe in this section.
We first summarize the methods used in the first-principles modelling of \ce{Li2NaN} in \cref{SM:Subsec:tb_model:dft}, before presenting a symmetry-consistent scheme for detaching the four most relevant bands near the Fermi level as well as our method of fitting the free parameters of the constructed TB model in \cref{SM:Subsec:tb_model:tb}.

\subsection{First-principles calculations}\label{SM:Subsec:tb_model:dft}

\begin{figure}[t!]
    \centering
    \includegraphics{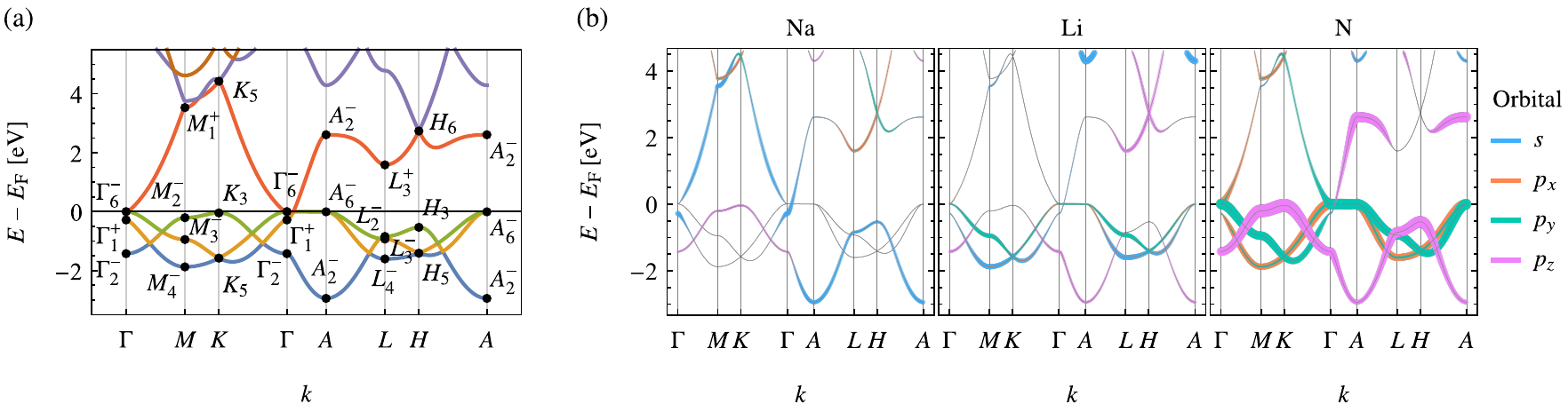}
    \caption{
        (a) Band structure of \ce{Li2NaN} along the lines of symmetry in an energy window around the Fermi level, obtained from first-principles calculations.
		The irreducible representations of all bands at the points of symmetry are shown.
		For brevity, we refer to the lowest-energy displayed band (blue) as `first', the next (yellow) as `second', and so forth.
        (b) Contribution of orbitals (colored overlays) to the band structure (gray lines) of \ce{Li2NaN}.
        The width of the colored overlays indicates the weight of the corresponding orbital.
    }
    \label{SM:fig:li2nan:dft}
\end{figure}

The tight-binding model is set up as an effective low-energy theory of the results of first-principles calculations.
Therefore, density functional theory (DFT) calculations are performed with the projected augmented wave (PAW) method as implemented in the Vienna ab initio simulation package (VASP)~\cite{Kresse:1996,Kresse:1999} with generalized gradient approximation (GGA) and PBE approximation ~\cite{Perdew:1996}.
We use a $\Gamma$-centered $10 \times 10 \times 8$ $k$-mesh.
Using plane-wave-based wave functions and space group operators generated by VASP, we calculate the traces of matrix representations to get the irreducible representations of the energy bands at high-symmetry points in the first Brillouin zone [cf.\ \cref{SM:fig:li2nan:dft}(a)] with the help of \texttt{irvsp}~\cite{Gao:2020}.
Using compatibility relations~\cite{Elcoro:2017} we then deduce that the irreducible representations of the lines of symmetry.
The contributions to energy bands from orbitals  [cf.\ \cref{SM:fig:li2nan:dft}(b)] are obtained by projecting wavefunctions to ion-centered spherical harmonic functions with different quantum numbers.
In the following we will number the bands starting from the yellow band in \cref{SM:fig:li2nan:dft}(a), which we will refer to as the `first' band, even though the first-principles results contain an additional band at even lower energy (approximately $-10\nunit{eV}$).

Furthermore, we rebuild the crystal structures for compressive and tensile strain by reducing and enlarging the $y$-component of each lattice vector.
Surface states are then calculated using the \texttt{WannierTools} package~\cite{Wu:2017} which is based on the maximally localized Wannier-function tight-binding model~\cite{Marzari:1997} constructed by the \texttt{Wannier90} package~\cite{Pizzi:2020}.
We use the sodium $s$-orbital, and nitrogen $p$-orbitals as initial wave functions for the localization.

\subsection{Effective low-energy tight-binding model}\label{SM:Subsec:tb_model:tb}

\begin{table}[bt!]
    \centering
    \begin{tabular}{x{1.7cm}x{1.7cm}x{1.7cm}x{1.7cm}x{1.7cm}}
    \hline\hline
        Element	& Orbital		& Position  	& Wyckoff pos.      & IR	            \\
		\hline\addlinespace
		\ce{Na}	& $s$			& $(0,0,1/2)$	& $b$				& $A_{1\mathrm{g}}$  \\
		\ce{N}	& $(p_x,p_y)$   & $(0,0,0)$		& $a$				& $E_{1\mathrm{u}}$  \\
		\ce{N}	& $p_z$         & $(0,0,0)$		& $a$				& $A_{2\mathrm{u}}$  \\
	\hline\hline
    \end{tabular}
    \caption{
        Orbitals used in the construction of the low-energy tight-binding model for \ce{Li2NaN}.
        The position is expressed in fractions of the primitive lattice vectors in \cref{SM:eq:lattice_vectors}.
        The last two columns give the corresponding Wyckoff position and irreducible representation (IR) of the site-symmetry group.
    }
    \label{SM:tab:li2nan:orbitals}
\end{table}

To construct an effective low-energy TB model for \ce{Li2NaN}, we need to choose a minimal set of orbitals.
From the first-principles data [\cref{SM:fig:li2nan:dft}(b)] we deduce that the band structure close to the Fermi level is dominated by four orbitals, namely sodium (\ce{Na}) $s$, and nitrogen (\ce{N}) $p_{x,y,z}$ [\cref{SM:tab:li2nan:orbitals}]. 
The compound contains one of each per the primitive unit cell, suggesting it is possible to construct a four-band model that reproduces the low-energy properties of \ce{Li2NaN}.
Note, however, that in the first-principles data the fourth band intersects with the fifth band multiple times, in particular leading to two-dimensional (2D) irreducible representations (IRs) at $K$ and $H$ [\cref{SM:fig:li2nan:dft}(a)].
Furthermore, other orbitals become dominant at these locations, notably the lithium (\ce{Li}) $p_z$ along $LH$ and $HA$ [\cref{SM:fig:li2nan:dft}(b)].
Thus, we need to artificially detach the fourth and fifth band, which prevents the resulting TB model from reproducing the band structure near those crossings.
Nevertheless, as these degenracies lie at energies far from the Fermi level (3~to 4~eV), this mismatch is not relevant for the low-energy properties.

We verify that such a detachment of bands is consistent with symmetries using the machinery of topological quantum chemistry (TQC)~\cite{Bradlyn:2017,Kruthoff:2017,Po:2017}.
Using the \texttt{WYCKPOS} and \texttt{POINT} applications on the Bilbao crystallographic server (BCS)~\cite{Aroyo:2006a,Aroyo:2006b} we determine the Wyckoff letters and irreducible representations (IRs) of the four orbitals in the model [\cref{SM:tab:li2nan:orbitals}].
These serve as an input for the \texttt{BANDREP} application~\cite{Bradlyn:2017,Elcoro:2017} on the BCS which calculates the elementary band representations (EBRs) induced from the IRs of the site-symmetry group of the Wyckoff positions [\cref{SM:tab:li2nan:bandreps}].
Matching them to the IRs in \cref{SM:fig:li2nan:dft}(a), we notice that not only the ones at the Dirac points at $K$ and $H$ need to be replaced, but also the one at $L$, namely
\begin{equation}
    K_5\to K_1,\quad H_6\to H_3,\quad L_3^+\to L_2^-.
\end{equation}
Due to band inversions, these IRs appear in the band structure at somewhat higher energies ($K_1$ is the eighth band at $E\approx 6.5\,\textrm{eV}$, $H_3$ is the ninth band at $E\approx 8.7\,\textrm{eV}$, and $L_2^-$ is the fifth band at $E\approx 5.0\,\textrm{eV}$).
Finally, we need to check the compatibility relations for the modified IRs of the detached fourth band to ensure that the observed band connectivity is compatible with the IRs.
Using the \texttt{COMPATIBILITY RELATIONS} application~\cite{Elcoro:2017} on the BCS we find that both $M_1^+ \to T_1 \to K_1$ and $A_2^- \to R_3 \to L_2^- \to S_3 \to H_3 \to Q_3 \to A_2^-$ are compatible.

\begin{table}[t!]
    \centering
    {\renewcommand{\arraystretch}{1.2}\begin{tabular}{x{2.2cm}x{2.2cm}x{2.2cm}x{2.2cm}}
    \hline\hline
		Wyckoff pos.	& $1a$ $(6/mmm)$				& $1a$ $(6/mmm)$				& $1b$ $(6/mmm)$				\\
		Band rep.  		& $E_{1\mathrm{u}}\uparrow G(2)$& $A_{2\mathrm{u}}\uparrow G(1)$& $A_{1\mathrm{g}}\uparrow G(1)$\\
	    \hline\addlinespace
		$\Gamma$		& $\Gamma_6^-(2)$				& $\Gamma_2^-(1)$				& $\Gamma_1^+(1)$				\\
		$M$				& $M_3^-(1)\oplus M_4^-(1)$		& $M_2^-(1)$					& $M_1^+(1)$					\\
		$K$				& $K_5(2)$						& $K_3(1)$						& $K_1(1)$						\\
		$A$				& $A_6^-(2)$					& $A_2^-(1)$					& $A_2^-(1)$					\\
		$L$				& $L_3^-(1)\oplus L_4^-(1)$		& $L_2^-(1)$					& $L_2^-(1)$					\\
		$H$				& $H_5(2)$						& $H_3(1)$						& $H_3(1)$						\\
		\hline\hline
    \end{tabular}}
    \caption{
        Elementary band representations (EBRs) induced from the irreducible representations of the site-symmetry group of the Wyckoff positions for the orbitals given in \cref{SM:tab:li2nan:orbitals}.
    }
    \label{SM:tab:li2nan:bandreps}
\end{table}

Having identified the relevant orbitals and the parts of the band-structure to be reproduced, we construct a symmetry-consistent TB model.
Given the orbitals and the isogonal point group $D_{6h}$ we determine the symmetry-restricted family of Bloch Hamiltonians using the \texttt{qsymm} \Python{} package~\cite{Varjas:2018}.
We choose primitive lattice vectors
\begin{equation}
	\vec{t}_1 = \mat{0\\-a\\0}, \quad \vec{t}_2 = \mat{{a\sqrt{3}}/{2}\\{a}/{2}\\0}, \quad \vec{t}_3 = \mat{0\\0\\c}
	\label{SM:eq:lattice_vectors}
\end{equation}
with $a$ and $c$ being the in-plane and out-of-plane lattice constants, respectively.
We further fix the Hilbert-space basis $(\mathrm{N}p_x,\mathrm{N}p_y,\mathrm{N}p_z,\mathrm{Na}s)$, i.e.\ in terms of the IRs $(E_{1\mathrm{u}}(2),A_{2\mathrm{u}}(1),A_{1\mathrm{g}}(1))$.

As the input to the \texttt{qsymm} package, we provide (\emph{i}) the set of generators of $D_{6h}$, including the matrix corepresentations which are determined by the IRs of the orbitals [listed in \cref{SM:tab:li2nan:tbsym}(left)], and (\emph{ii}) the considered hopping vectors [listed in \cref{SM:tab:li2nan:hoppings}(right)] between nearby atoms.
We only consider hoppings up to nearest neighbors in terms of unit cells, which results in six independent terms (all symmetry-related hopping terms are taken into account implicitly).
The relevant \Python{} script as well as its output can be found in the \SuppData{}.
We note that the resulting family of Bloch Hamiltonians does not consist of real matrices.
However, because we need the Hamiltonian to be real when computing the quaternion invariant, we rotate the basis such that $\mcP\mcT=\id\mcK$ using the unitary transformation $U=\diag(1,1,1,\imi)$.

\begin{table}[t!]
    \centering
    \begin{tabular}{cccc}
    \hline\hline
		Generator         	& Action on position		& Matrix corepresentation		\\
		\hline\addextralinespace{2}
		$C_6^+$				& $\mat{\tfrac{1}{2}&-\tfrac{\sqrt{3}}{2}&0\\\tfrac{\sqrt{3}}{2}&\tfrac{1}{2}&0\\0&0&1}$ & $\mat{\tfrac{1}{2}&-\tfrac{\sqrt{3}}{2}&0&0\\\tfrac{\sqrt{3}}{2}&\tfrac{1}{2}&0&0\\0&0&1&0\\0&0&0&1}$ \\
		\addextralinespace{2}
		$C_{21}'$			& $\mat{1&0&0\\0&-1&0\\0&0&-1}$ & $\mat{1&0&0&0\\0&-1&0&0\\0&0&-1&0\\0&0&0&1}$ \\
		\addextralinespace{2}
		$\mcT$				& $\mat{1&0&0\\0&1&0\\0&0&1}$ & $\mat{1&0&0&0\\0&1&0&0\\0&0&1&0\\0&0&0&1}$ \\
		\addextralinespace{2}
		$\mcP$					& $\mat{-1&0&0\\0&-1&0\\0&0&-1}$ & $\mat{-1&0&0&0\\0&-1&0&0\\0&0&-1&0\\0&0&0&1}$ \\
	\hline\hline
    \end{tabular}
    \hspace{2cm}
    {\renewcommand{\arraystretch}{1.6}\begin{tabular}{cccc}
    \hline\hline
		Atom 1		& Atom 2	& Hopping vector					\\
		\hline\addlinespace
		\ce{N}		& \ce{N}	& $\vec{t}_1$						\\
		\ce{N}		& \ce{N}	& $\vec{t}_3$						\\
		\ce{Na}		& \ce{Na}	& $\vec{t}_1$						\\
		\ce{Na}		& \ce{Na}	& $\vec{t}_3$						\\
		\ce{N}		& \ce{Na}	& $\tfrac{1}{2}\vec{t}_3$			\\
		\ce{N}		& \ce{Na}	& $\vec{t}_1+\tfrac{1}{2}\vec{t}_3$	\\
		\hline\hline
    \end{tabular}}
    \caption{
        (left) Generators of the isogonal point group $D_{6h}$ including their action on position and the matrix corepresentation in the Hilbert space spanned by the orbital basis $(\mathrm{N}p_x,\mathrm{N}p_y,\mathrm{N}p_z,\mathrm{Na}s)$ [\cref{SM:tab:li2nan:orbitals}]. (right) Hopping terms between the atoms listed in~\cref{SM:tab:li2nan:orbitals}.
    }
    \label{SM:tab:li2nan:tbsym}\label{SM:tab:li2nan:hoppings}
\end{table}

\begin{figure}[tb!]
    \centering
    \includegraphics{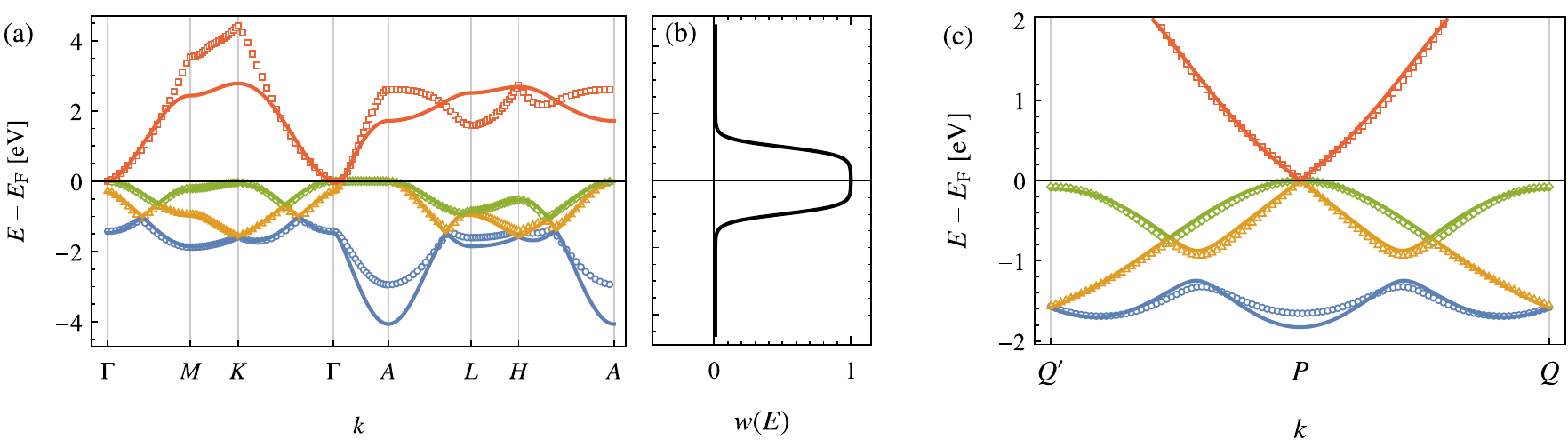}
    \caption{
        Band structure of \ce{Li2NaN} (a) along the lines of symmetry and (c) near the triple point.
        First-principles data (only a fraction of all the points is shown for better visibility) is indicated by points and the fitted tight-binding model by solid lines. 
        (b) The fitting weight function given in \cref{SM:eq:weight-function}.
        For the description of the $Q'PQ$ path in panel (c) see the paragraph preceding \cref{SM:eq:li2nan:strain}.
        The coloring of the energy bands follows their labelling according to increasing energy.
        }
    \label{SM:fig:li2nan:bs}
\end{figure}

We arrive at a 13-parameter family of Bloch Hamiltonians, which we now fit to the first-principles band structure along the lines of symmetry.
We cannot expect the model to fit well everywhere, because it contains a reduced number of orbitals, and because of the symmetry-consistent replacement of IRs at points $K$, $H$ and $L$, as argued above.
In order to capture the relevant features, i.e.\ the energies close to the Fermi level and the band connectivity, we proceed as follows:
First, we set the energy of the 2D IR along $\Gamma A$ exactly to zero, which is almost exact according to the first-principles data.
Then, we determine initial values for ten out of the remaining eleven parameters by diagonalizing the eleven-parameter family of Hamiltonians on lines of symmetry and using the dimension of the IRs to match the resulting expressions for eigenvalues to the numerically obtained band structure.
Finally, we fit the eleven parameters not fixed in the first step using a weighted least-square algorithm with the weight function 
    \begin{equation}
    	w(E) = \frac{w_0+\frac{1}{2}\coth\left(\lambda\frac{\Delta E}{2}\right)\left[\tanh\left(\lambda\left(\frac{\Delta E}{2}+E\right)\right)+\tanh\left(\lambda\left(\frac{\Delta E}{2}-E\right)\right)\right]}{1+w_0},\label{SM:eq:weight-function}
    \end{equation}
where $\lambda=4\nunit{eV^{-1}}$, $\Delta E = 2\nunit{eV}$, $w_0=0.01$ and $E$ is the band energy according to the first-principles calculations.
The choice of weight function in \cref{SM:eq:weight-function}, illustrated in \cref{SM:fig:li2nan:bs}(b), ensures that the fitted model reproduces features within an approximately $\Delta E/2 = 1\,\textrm{eV}$ wide energy window around the Fermi level more faithfully than those farther away.
We provide an implementation of the fitting procedure in \Mathematica{} as well as our results in the \SuppData{}.

The band structure of the resulting TB model along the lines of symmetry is compared to the first-principles data in \cref{SM:fig:li2nan:bs}(a).
We observe that the fit is good close to the Fermi level but diminishes for energies farther away.
Except for the fourth band, the band connectivity and overall behaviour of the bands is reproduced, as expected.
\Cref{SM:fig:li2nan:bs}(c) shows the band structure along the line $Q'PQ$ with $P$ being the position of the triple point and $Q$ a point on the BZ boundary directly above $K$ at the same $k_z$ as $P$.
We thus expect the TB model to reproduce the band structure in the energy window $[-2,2]\nunit{eV}$ very well even outside the lines of symmetry and in particular near the triple point.

To convert the TPs into multiband nodal links, the $C_{6v}$ symmetry needs to be broken.
In the TB model this is modelled by adding the leading $C_{6v}$-breaking term,
\begin{equation}
	\Ham_\text{strain}(\vec{k}) = s\mat{1&&&\\&-1&&\\&&0&\\&&&0},
	\label{SM:eq:li2nan:strain}
\end{equation}
which lifts the degeneracy along the $\Gamma A$ line. 
This term still satisfies the $C_{2v}$ symmetry and thus corresponds to the application of uniaxial strain.

\begin{figure}[t]
    \centering
    \includegraphics{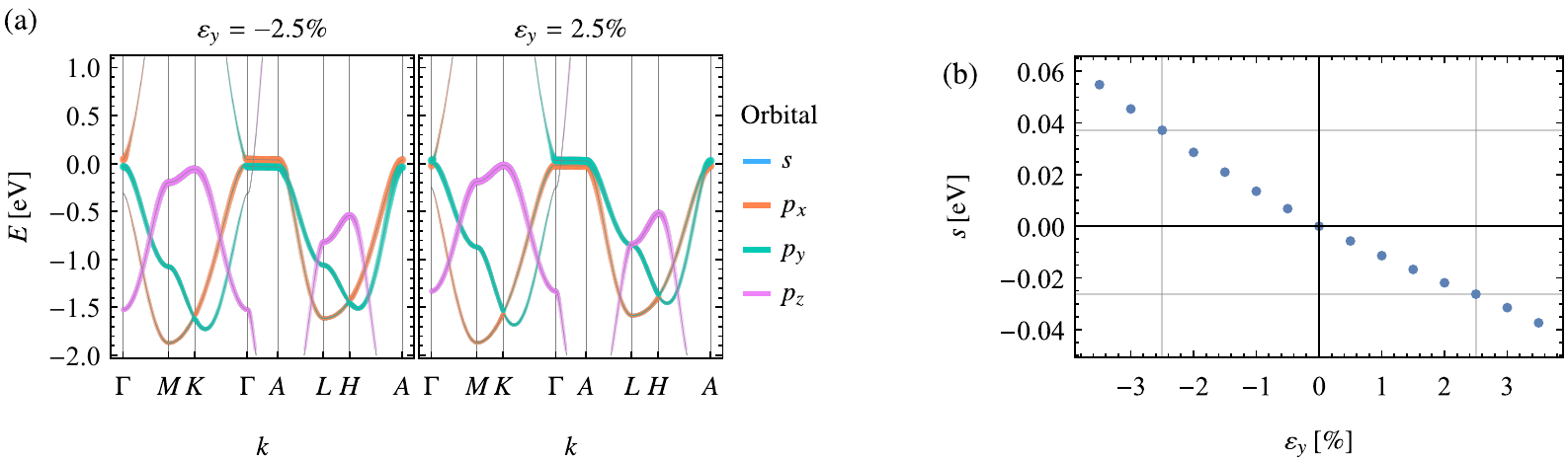}
    \caption{
        First-principles data showing the effect of uniaxial strain $\varepsilon_y$.
        (a) Band structure and contribution of the \ce{N} orbitals (colored overlays) for compression and expansion in the $y$-direction. The energy of the band dominated by $p_x$ decreases for $\varepsilon_y>0$ and increases for $\varepsilon_y<0$.
        (b) The extracted dependence of the energy splitting parameter $s$ on the strain $\varepsilon_y$.
    }
    \label{SM:fig:li2nan:strain_param}
\end{figure}

In the first-principles calculations, we model uniaxial strain by deforming the lattice constants, such that a $C_{2v}$ symmetry remains, see \cref{SM:Subsec:tb_model:dft} for more details on the numerics.
In particular, we multiply the $y$-component of all lattice vectors by a factor of $1+\varepsilon_y$ with experimentally reasonable values of $\varepsilon_y$ in the range from $-3.5\%$ to $3.5\%$.
The dominant effect is the splitting of the 2D IR along $\Gamma A$, which can be related to the strain parameter $s$ from \cref{SM:eq:li2nan:strain} by
\begin{equation}
    2s \;\widehat{=}\; E_{\ce{N}p_x}(\Gamma A) - E_{\ce{N}p_y}(\Gamma A).
\end{equation}
Extracting the value of $s$ for different $\varepsilon_y$ we find that $\varepsilon_y>0$ corresponds to $s<0$ [\cref{SM:fig:li2nan:strain_param}(a)] and can establish the relationship $s(\varepsilon_y)$ [\cref{SM:fig:li2nan:strain_param}(b)].
For example, $s=-0.03\nunit{eV}$ corresponds to roughly $3\%$ tensile strain in the $y$-direction.

\section{Finding nodal lines based on a tight-binding model}\label{SM:Sec:NL_finding}

\begin{figure}[tb!]
    \centering
    \includegraphics{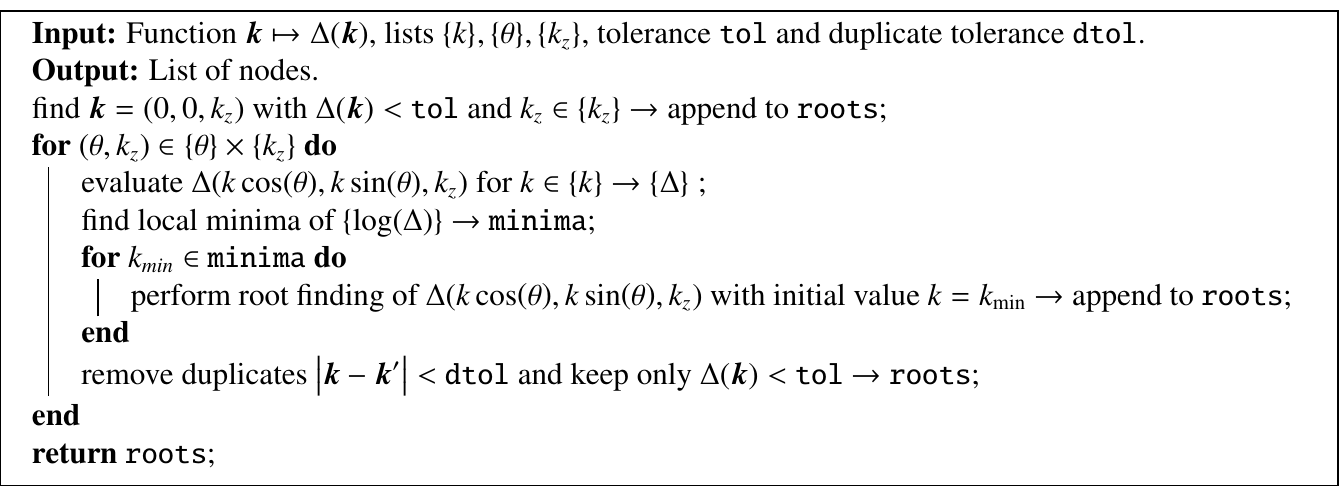}
    \caption{
        Algorithm for finding nodal lines based on roots of the discriminant $\Delta$ of the characteristic polynomial of the Bloch Hamiltonian.
    }
    \label{SM:fig:alg:nl_finding}
\end{figure}

Based on the ideas described in our follow-up paper~\cite{Lenggenhager:2022:TPClassif}, we have developed a simple partially adaptive numerical algorithm for finding nodal lines given a TB model.
It is based on the fact that the discriminant $\Delta(\vec{k})$ of the characteristic polynomial of the Hamiltonian $\Ham(\vec{k})$ vanishes if and only if there is a node at $\vec{k}$ (corresponding to a multiple root of the characteristic polynomial).
Because the Hamiltonian is Hermitian, the roots of the characteristic polynomial are always real, which implies that $\Delta\geq 0$.
Thus, the roots of $\Delta$ are simultaneously local minima.
It is convenient to go to cylindrical coordinates
\begin{equation}
	\vec{k} = \mat{k\cos(\theta)\\k\sin(\theta)\\k_z}
\end{equation}
and consider $k=0$ and $k\neq 0$ separately.
The first case can be checked explicitly for roots, while we handle the second case using a grid of $(\theta, k_z)$ points.
For each $(\theta, k_z)$ we sample $k$, detect minima and perform root finding around those.
At the end, roots are selected according to a specified tolerance, duplicates are removed and the energy gap is determined by computing the eigenenergies.
The most important parts of the algorithm are summarized in \cref{SM:fig:alg:nl_finding} in pseudocode and our implementation in \Mathematica{} and the results can be accessed in the \SuppData{}.

The algorithm has one limitation:
NLs confined to planes with constant $\theta$ and/or $k_z$ are not detected if the corresponding values of $\theta$ and $k_z$ are not in the initial grid.
For mirror-symmetry protected NLs this can be easily alleviated by making sure that all candidate planes (according to the symmetries) are included in the $(\theta, k_z)$ grid, while other NLs do not generally lie inside such a plane, such that they will be detected for some values of $k_z$ at least.
Alternatively, one could add additional adaptive loops to take care of that.
We have chosen the former way to deal with that limitation and found it sufficient for our purposes.

\section{Bulk-boundary correspondence for quaternion charges in \texorpdfstring{\NoCaseChange{\ce{Li2NaN}}}{Li2NaN} without and with strain}\label{SM:Sec:bulk_boundary}

In this section we test the bulk-boundary correspondence for the generalized quaternion charge as proposed by Ref.~\onlinecite{Wu:2019}.
To do so, we first review in \cref{SM:Subsec:bulk_boundary:quaternions} the main concepts relevant for understanding and computing the generalized quaternion charge, following closely the discussion in Ref.~\onlinecite{Wu:2019}.
Based on that discussion, we present in \cref{SM:fig:alg:quaternion_inv} an algorithm for the computation of the quaternion charge.
In \cref{SM:Subsec:bulk_boundary:surface_spectrum} we then discuss the surface spectrum of \ce{Li2NaN} without and with strain, and relate it to the bulk quaternion charges.
We observe an agreement with the anticipated bulk-boundary correspondence after the surface states that hybridize with the bulk states are explicitly included.

\subsection{Review of the generalized quaternion charge}\label{SM:Subsec:bulk_boundary:quaternions}

Consider an $N\times N$ real Bloch Hamiltonian in one dimension for generic $N\geq 2$.
By the spectral theorem, it can be written as
\begin{equation}
	\Ham(k) = \mathsf{u}(k)\mathcal{E}(k)\mathsf{u}(k)^T,
\end{equation}
where $\mathsf{u}(k)$ is the matrix with columns being the (ordered, real, orthonormal) eigenstates $\vec{u}_j(k)$ and $\mathcal{E}(k)$ the diagonal matrix of (ascending) eigenenergies of $\Ham(k)$.
The matrix $\mathsf{u}(k)$ is an orthogonal real matrix and thus defines an $N$-frame, the \emph{eigenframe}.
Note that the band energies $\mathcal{E}_j(k)$ can be continuously adjusted (as long as we do not produce a degeneracy) without affecting the band topology.
Thus, as long as there are no degeneracies, the band topology of the Hamiltonian can be completely captured by the eigenframe $\mathsf{u}(k)$.
Crucially, the frame has a remaining gauge freedom that manifests as the flipping of the sign of some of the vectors $\vec{u}_j$ of the frame.
This is given by the $N$-dimensional point group $\PN_{Nh}=\O(1)^N=\mathbb{Z}_2^N$ generated by $N$ mutually perpendicular mirror symmetries.
We can fix the handedness of the frame by choosing a gauge where $\mathsf{u}(k)\in\SO(N)$, which then reduces $\PN_{Nh}$ to its special component $\PN_N$.
Consequently, the space of (spectrally normalized) Hamiltonians that we should consider~\cite{Wu:2019} is
\begin{equation}
	M_N = \O(N)/\PN_{Nh} = \SO(N)/\PN_N = \Spin(N)/\overline{\PN}_N,\label{eqn:groups-quotients}
\end{equation}
where we first restricted to the special components and then lifted each group to its respective double cover.
We explain further below that the fundmamental group of $M_N$, which coincides with the discrete group $\overline{\PN}_N$, captures the topology of NLs in spinless $\mcP\mcT$-symmetric models.

We first briefly characterize the groups and lifts appearing in \cref{eqn:groups-quotients}.
The lift of an element of $\SO(N)$ close to the identity to $\Spin(N)$ is performed by working in the corresponding algebras $\so(N)$ and $\spin(N)$, where the lift can be simply performed on the level of the basis elements:
\begin{equation}
	\sum_{i<j}\alpha_{ij}L_{ij} \mapsto \sum_{i<j}\alpha_{ij}t_{ij}.
\end{equation}
Here, $\{L_{ij}\}_{i<j}$ and $\{t_{ij}\}_{i<j}$ are bases of $\so(N)$ and $\spin(N)$, respectively, and we choose
\begin{IEEEeqnarray}{rCl}
    (L_{ij})_{ab} &=& -\delta_{ia}\delta_{jb}+\delta_{ib}\delta_{ja},\\
    t_{ij} &=& -\frac{1}{4}\comm{\Gamma_i}{\Gamma_j},
\end{IEEEeqnarray}
where the $\{\Gamma_i\}$ are Gamma matrices in $N$ dimensions.
Furthermore, $\PN_N$ is a subgroup of $\O(N)$, and we can obtain the generators $\{e_1,e_2,\dotsc,e_{N-1}\}$ of $\overline{\PN}_N$ from the generators $\big\{\e{\pi L_{1j}}\big\}_{j=2}^N$ of $\PN_N$:
\begin{equation}
	e_{j-1} = \e{\pi t_{1j}} = \frac{1}{2}\comm{\Gamma_j}{\Gamma_1} = 2t_{1j}
\end{equation}
for $2\leq j\leq N-1$. From its definition we can read off that $e_{j-1}$ encodes a $\pi$-rotation in the $(1,j)$-plane (i.e.~a $\pi$-twist of the first and $j^\textrm{th}$ band).
An alternative set of generators is provided by
\begin{equation}
	g_j \equiv
	\begin{cases}
		e_1,&j=1\\
		e_{j-1}e_j,&j\geq 2
	\end{cases},
	\label{SM:eq:Q_generators}
\end{equation}
which corresponds to a $\pi$-rotation in the $(j,j+1)$-plane [i.e.~the band inversion of the $j^\textrm{th}$ and $(j+1)^\textrm{th}$ band].
The elements of $\overline{\PN}_N$ are then given by all possible products of elements of either set of generators together with the identity.
We call this group the \emph{generalized quaternion group} $\Q_N$ and denote its $2^N$ elements by
\begin{equation}
	\unarypm 1,\;\unarypm\mathfrak{q}_1,\;\unarypm\mathfrak{q}_2,\;\dotsc,\;\unarypm\mathfrak{q}_{2^{N-1}-1},
\end{equation}
where the corresponding matrix representations $q_i$ of the elements $\mathfrak{q}_i$ are the products of generators:
\begin{equation}
	q_1 = g_1,\quad
	q_2 = g_2,\quad
	\dotsc,\quad
	q_{N-1} = g_{N-1},\quad
	q_{N} = g_1g_2,\quad
	q_{N+1} = g_1g_3,\quad
	\dotsc,\quad
	q_{2^{N-1}-1}=g_1g_2\dotsb g_{N-1}.\label{eqn:quat-matrices}
\end{equation}

\begin{figure}[bt!]
    \centering
    \includegraphics{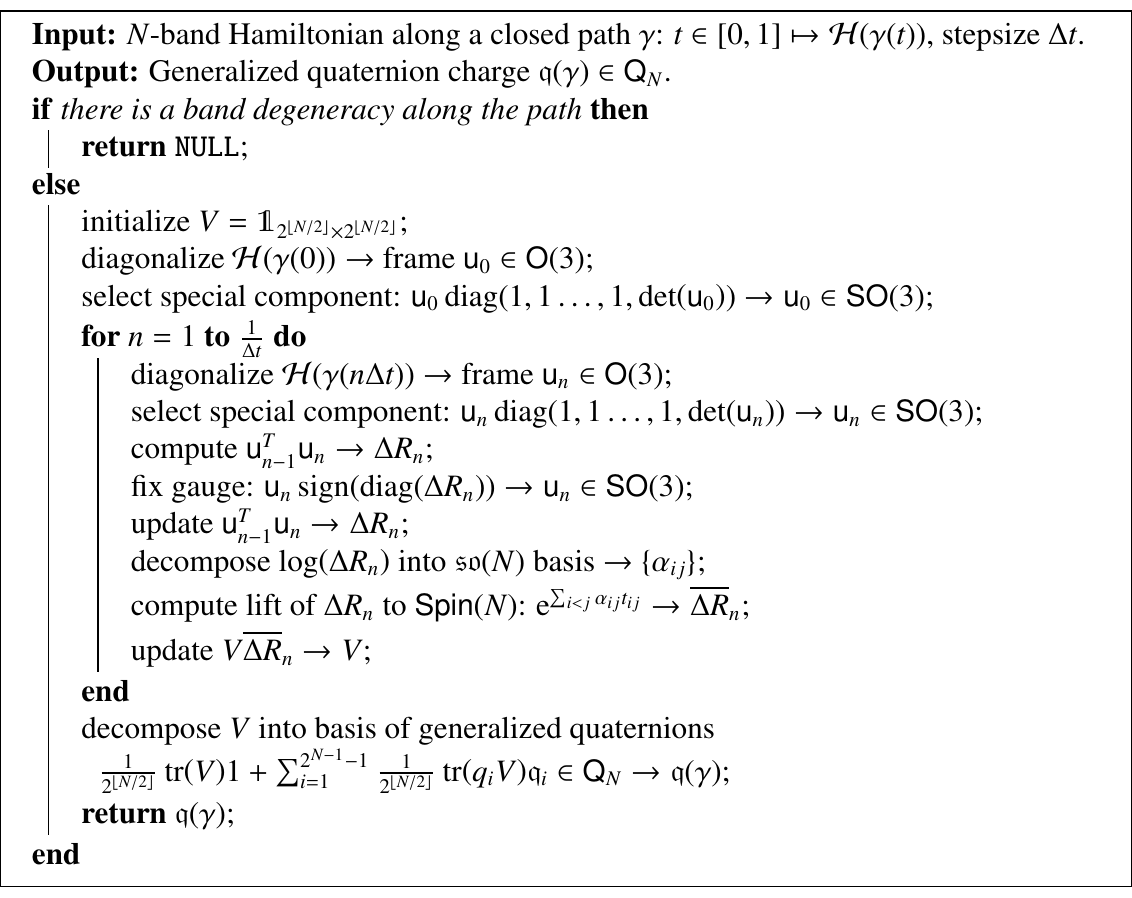}
    \caption{
        Computation of the generalized quaternion charge $\mathfrak{q}(\gamma)\in\Q_N$ for an $N$-band Hamiltonian $\Ham(\vec{k})$ along a closed path $\vec{k}=\gamma(t)$, based on the discussion in Ref.~\onlinecite{Wu:2019}.
        A \emph{frame} is defined to be the matrix with columns being the ordered real orthonormal eigenstates in order of ascending eigenenergies and $\sign(\diag(\Delta R_n))$ is the diagonal matrix consisting of the signs the diagonal entries of $\Delta R_n$.
    }
    \label{SM:fig:alg:quaternion_inv}
\end{figure}

We now turn attention to the topological invariants in the momentum space.
Closed paths $\gamma$ in momentum space with non-degenerate spectrum at each point $k\in\gamma$ are characterized by the first homotopy group $\pi_1(M_N)$, which is found to be~\cite{Wu:2019}
\begin{IEEEeqnarray}{rCl}
    \pi_1(M_N) &=& \overline{\PN}_N.
\end{IEEEeqnarray}
This implies the existence of an invariant $\mathfrak{q}(\gamma)\in\overline{\PN}_N$, called \emph{generalized quaternion charge} by Ref.~\onlinecite{Wu:2019}.
Its numerical computation is based on tracking the rotation of the eigenframe lifted to $\Spin(N)$ along the closed path $\gamma$.
The total spin-rotation along $\gamma$ can be decomposed into a linear combination of the matrices in \cref{eqn:quat-matrices}, and if there are no band degeneracies along $\gamma$, the total rotation itself is (within numerical precision) equal to one of these matrices, i.e. it is an element of $\Q_N$.
\Cref{SM:fig:alg:quaternion_inv} shows this algorithm in pseudocode.
We make our implementation in \Mathematica{} available in the \SuppData{}.

Due to its non-Abelian character, the quaternion charge is only well-defined up to conjugation by elements in $\Q_N$, such that the topology is classified by the equivalences classes of $\Q_N$.
They are~\cite{Wu:2019}
\begin{itemize}
	\item $\{\pm g_1^{n_1}g_2^{n_2}\cdots g_{N-1}^{n_{N-1}}\}$, $n_i\in\{0,1\}$ with the exceptions
	\item $\{+1\}$, $\{-1\}$ and
	\item if $N$ is even also $\{+g_1g_3\cdots g_{N-1}\}$, $\{-g_1g_3\cdots g_{N-1}\}$,
\end{itemize}
which follow from the commutation relations between the generators.
The fact that $\pm 1$ are in distinct conjugacy classes for any $N$ results in the paths that are classified by these charges to be stably topologically inequivalent, while $\pm g_1g_3\cdots g_{N-1}$ result in an unstable topological inequivalence (since adding an additional band brings them into a single conjugacy class).

The set of generators given in \cref{SM:eq:Q_generators} provides a useful interpretation of the quaternion charge in terms of its factors $g_j$.
Since $g_j$ corresponds to a $\pi$-rotation in the $(j,j+1)$-plane, it suggests 
a band inversion between bands $(j,j+1)$~\cite{Wu:2019,Tiwari:2020} with associated $\pi$ Berry phases on those two bands.
Consequently, if the quaternion charge is computed on a closed loop in $\vec{k}$-space that does not wind around the BZ torus, then each factor $g_j$ indicates that a band node (in 2D) or NL (in 3D) between bands $(j,j+1)$ is enclosed~\cite{Wu:2019}.
Equivalently, we consider the Hamiltonian along the closed loop as a 1D system.
Then, deforming it to a constant Hamiltonian requires a closing of the gap between bands $(j,j+1)$ for each factor $g_j$.
We make use of this interpretation in the following study of the bulk-boundary correspondence.

\subsection{Surface Spectrum \texorpdfstring{\NoCaseChange{\ce{Li2NaN}}}{Li2NaN} without and with strain}\label{SM:Subsec:bulk_boundary:surface_spectrum}

\begin{figure*}[b!]
    \centering
    \includegraphics{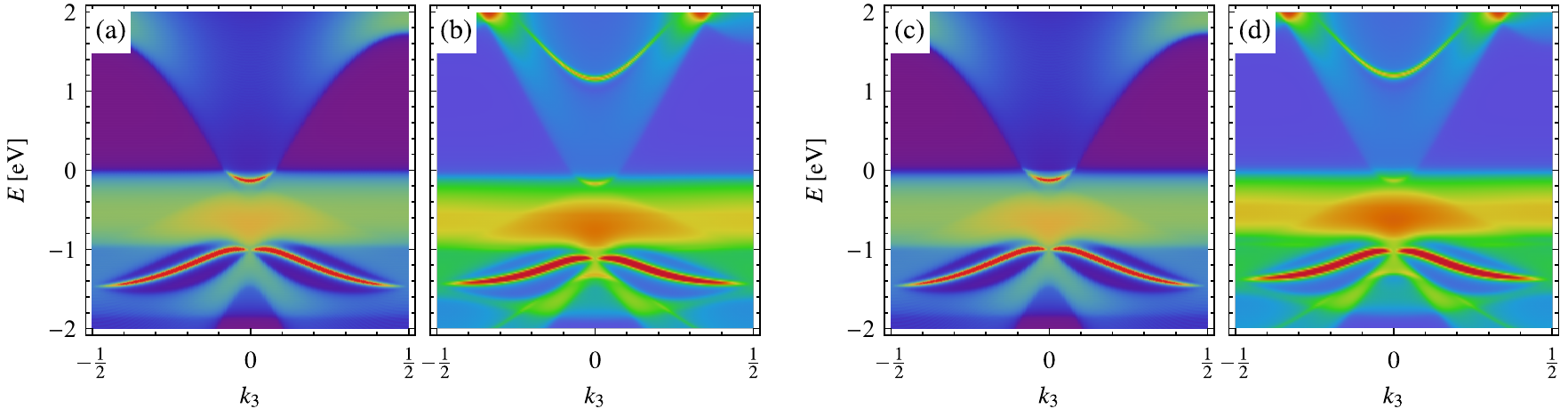}
    \caption{
        Surface spectral function of \ce{Li2NaN} (a,b) without strain and (c,d) with $3\%$ tensile strain (corresponding to $s\approx -0.03\nunit{eV}$), which was obtained from (a,c) the tight-binding model with $N=100$ layers [cf.\ \cref{SM:Subsec:tb_model:tb}] and (b,d) first-principles calculations [cf.\ \cref{SM:Subsec:tb_model:dft}]. 
        Different relative color scales are used in the comparisons.
        Note that the surface band at energy $E>1\,\textrm{eV}$ visible in the first-principles data (b,d) is not reproduced in our TB model (a,c). This difference arises due to the enforced detachment of the fourth band from the higher-energy ones.
    }
    \label{SM:fig:li2nan:surf_spec_dft}
\end{figure*}

To analyze the surface properties we consider a slab geometry of the TB model $\Ham(\vec{k})$~\cite{SuppData} with termination at the surface indicated in \cref{fig:li2nan}(a) of the main text.
In that illustration, the atoms intersected by the red layer are \emph{included} in the slab, and the purple arrow indicates the surface normal pointing \emph{outside} from the sample.
Thus, we assume a finite-size system with open boundary conditions in the $\vec{t}_1$ direction and an infinite size in the $\vec{t}_2,\vec{t}_3$ directions. The Bloch Hamiltonian for $N$ layers is
\begin{equation}
    \Ham(k_2,k_3)_{nn'} = \int_{-1/2}^{1/2}\dd{k_1}\Ham\left(\sum_{j=1}^3k_j\vec{G}_j\right)\e{2\pi\i(n-n')k_1},
    \label{SM:eq:H_slab}
\end{equation}
where $n\in\{1,2,\ldots,N\}$ is the layer index.
By diagonalizing $\Ham(k_2,k_3)$ at each point $(k_2,k_3)$ in the surface Brillouin zone (SBZ),
\begin{equation}
    \Ham(k_2,k_3)\ket{i} = E_i\ket{i},
\end{equation}
we obtain $4N$ eigenstates, and compute the surface spectral function
\begin{equation}
    A_\text{S}(E,k_2,k_3) = \sum_i\delta(E_i-E)\mel{i}{P_\text{surface}}{i},
    \label{SM:eq:surf_spec}
\end{equation}
where $P_\text{surface}$ is the projector onto the outermost surface layer $n=N$.
For our calculations we use $N=100$ and we replace the Dirac delta function by a Lorentzian with a half-width of $0.021\nunit{eV}$.
The results of the calculation for the TB model without and with strain are shown in Figs.~\ref{SM:fig:li2nan:surf_spec_dft}(a) and~\ref{SM:fig:li2nan:surf_spec_dft}(c), respectively.

We compare the surface spectral function obtained from the TB model to the predictions from the first-principles modelling [cf.\ \cref{SM:Subsec:tb_model:dft}].
While the constructed TB model does not include the lithium atoms, thus leading to a unique surface termination along the considered direction, the explicit presence of the lithium orbitals in the first-principles calculations enables terminations at various inequivalent layers.
By comparing the data, we find that the TB model reproduces the first-principles calculations [\cref{SM:fig:li2nan:surf_spec_dft}(b,d)] for the termination at the layer shown in \cref{fig:li2nan}(a) of the main text with the specifications detailed above \cref{SM:eq:H_slab}. For this termination, the outermost layer of nitrogen atoms is enclosed in a complete hexagonal cage of lithium atoms, just like the bulk nitrogen atoms.

The observed correspondence suggests that the basis orbitals of the constructed TB model effectively correspond to a mixture of the nitrogen $p_{x,y}$ orbitals hybridized with the lithium $p_{x,y}$ orbitals of the enclosing hexagonal cage [as is indeed visible in \cref{SM:fig:li2nan:dft}(c)].
Since the TB model implicitly assumes the same orbitals also at the boundary, the surface termination reproduced by the TB model must correspond to the case when the outermost nitrogen atoms are surrounded by unbroken hexagonal cages of lithium atoms, consistent with our findings.
One obvious difference between the predictions of the TB model [\cref{SM:fig:li2nan:surf_spec_dft}(a,c)] as compared to the first-principles data [\cref{SM:fig:li2nan:surf_spec_dft}(b,d)] is the absence of a surface band in energy range of $1$ to $2\,\textrm{eV}$, i.e.\ relatively far from the Fermi level. We expect this surface band to originate from a band inversion between the fourth and higher-energy bands, and that it is absent in the TB modelling due to the enforced detachment of those bands [cf.\ \cref{SM:Subsec:tb_model:tb}].

\begin{figure*}[t!]
    \centering
    \includegraphics{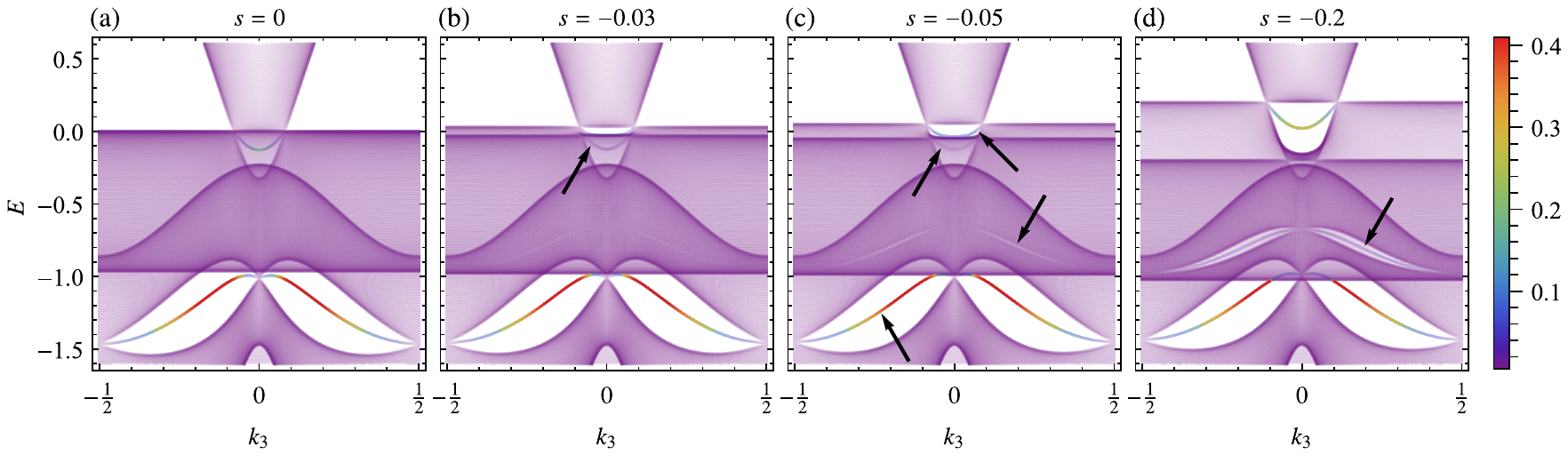}
    \caption{
    Band structure of the slab Hamiltonian of \ce{Li2NaN} defined in \cref{SM:eq:H_slab} for $N=200$ layers and $k_2=0$, plotted for increasing strain given by the parameter $s$ defined in \cref{SM:eq:li2nan:strain}.
    The color scale indicates the inverse participation ratio (IPR), \cref{SM:eq:IPR}, according to the legend on the right.
    In panel (a) the IPR reveals two surface bands, one in gap $1\--2$ and one in gap $2\--3$.
    In the strained case [panels (b--d)], not all surface bands are well visible, due to hybridization with the bulk states, whose degree depends on $s$.
    However, looking at panels (b--d) we can deduce that there are four surface bands [indicated by black arrows in panel (c)], even if they are not always visible [e.g.\ the upper surface state in gap $2\-- 3$ is best visible in panel (b) and the lower one in panel (d)].
    }
    \label{SM:fig:li2nan:slab:Ek3s}
\end{figure*}

\begin{figure*}[b!]
    \centering
    \includegraphics{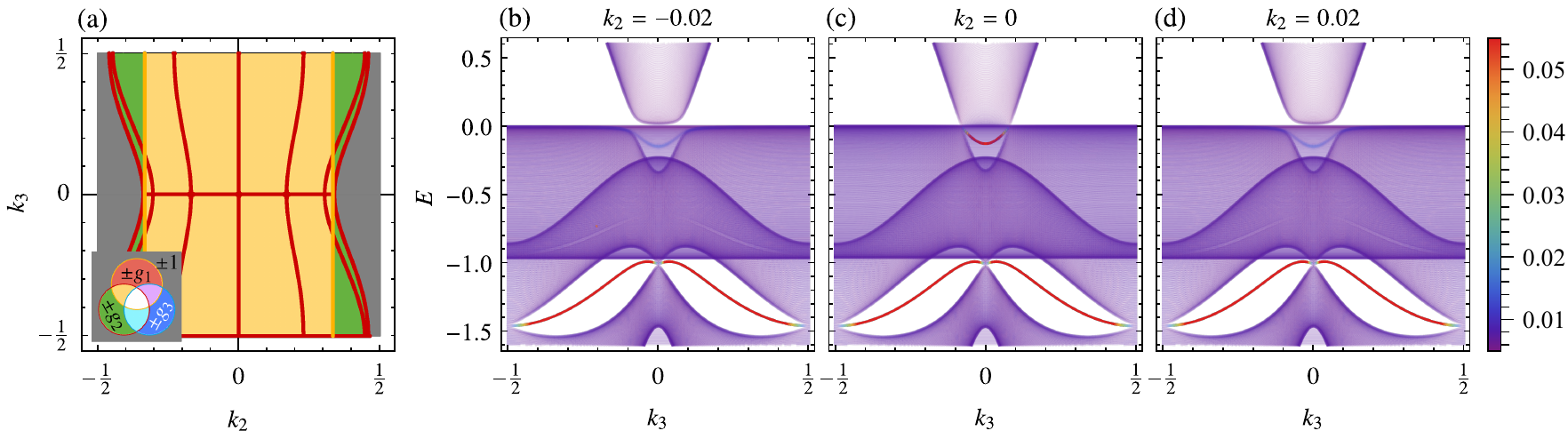}
    \caption{
    (a) Quaternion charge in the full surface Brillouin zone and the projection of bulk nodal lines for $s=0$. As indicated in the legend, we assign colors to the equivalence classes of the quaternion charge according to the representation in terms of the generators $g_1, g_2, g_3$, such that e.g.\ the overlap of red and green (yellow) corresponds to $\pm g_1g_2$.
    For illustration purposes we do not distinguish between elements of different signs even in the cases where they are in \emph{different} equivalence classes ($\pm 1$ and $\pm g_1g_3$).
    This does not impair the presentation of results, because the gray region coincidentally is always $+1$, while the distinction of $\pm g_1g_3$ is fragile under the addition of additional trivial bands.
    (b--d) Band structure of the unstrained slab Hamiltonian of \ce{Li2NaN} defined in \cref{SM:eq:H_slab} for $N=200$ layers and different values of $k_2$, corresponding to three cuts through the surface Brillouin zone.
    The color scale indicates the inverse participation ratio (IPR), \cref{SM:eq:IPR}, according to the legend on the right.
    The cuts at $k_2\neq 0$ clearly reveal that the two surface bands lie in gaps $1\--2$ and $2\--3$.
    }
    \label{SM:fig:li2nan:slab:Ek3k2_noStrain}
\end{figure*}

To better visualize the surface states, we now consider the full slab band structure for $N=200$ layers instead of just the surface spectral function, and we compute the inverse participation ratio (IPR) of each state.
The IPR of a (normalized) state vector $\psi$ is defined as
\begin{equation}
    \textrm{IPR}(\psi) = \sum_{n=1}^N p_n(\psi)^{2},
    \label{SM:eq:IPR}
\end{equation}
where the sum is over the layer index $n$, and $p_n(\psi)$ is the probability of an electron in state $\psi$ to be found in layer $n$ (in any of the four basis orbitals).
The IPR is a measure of localization: a state $\psi$ perfectly and uniformly localized on $m$ layers satisfies $\textrm{IPR}(\psi)=1/m$, which for a completely delocalized state becomes $1/N$.

In general, surface states can hybridize with bulk states that project to the same point in the SBZ and lie at the same energy.
However, as long as the hybridization is not too strong, they are still localized and are thus visible via the IPR.
To uncover \emph{all} surface states, we look at the evolution of the band structure (and IPR) when increasing the strain parameter $s$ from $0$ to $-0.2$ [\cref{SM:fig:li2nan:slab:Ek3s}] for $k_2=0$.
In the strained case [panels (b--d)] we identify a total of four surface bands: one formed by bands $1\--2$, two by $2\--3$ and one by $3\--4$.
For large enough $s$, the first and last surface band are located in bulk energy gaps, while the other two clearly hybridize with bulk states. In the unstrained case [panel (a)], only two surface bands are visible, one between bands $1\--2$, and one between bands $2\--3$.

\begin{figure}[t!]
    \centering
    \includegraphics{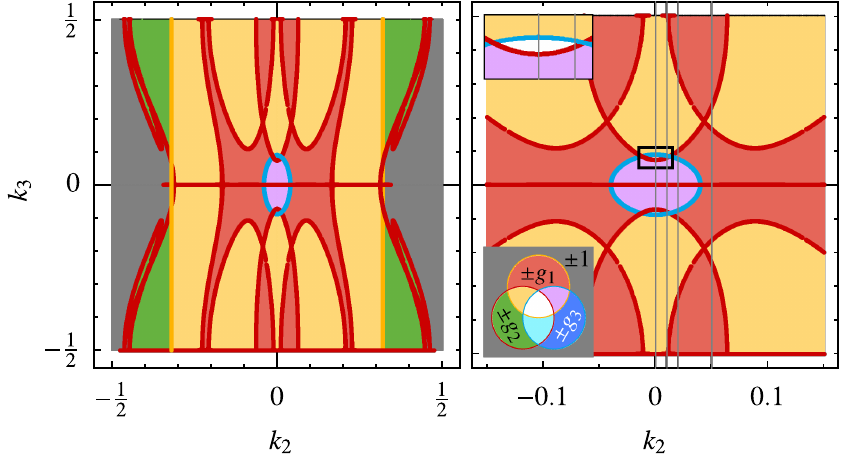}
    \caption{
        Quaternion charge and the projection of bulk nodal lines for $s=-0.05\nunit{eV}$.
        See caption to \cref{SM:fig:li2nan:slab:Ek3k2_noStrain}(a) for a description of the legend.
        The panel on the left shows the full surface Brillouin zone (SBZ), the panel on the right a close-up on $\abs{k_2}<0.15$, with the inset showing the details of the multiband nodal link.
        The four vertical gray lines in the right panel indicate the cuts for which the slab band structure and IPR are plotted in \cref{SM:fig:li2nan:slab:Ek3k2}.
    }
    \label{SM:fig:li2nan:SBZQcharge}
\end{figure}

\begin{figure*}[t!]
    \centering
    \includegraphics{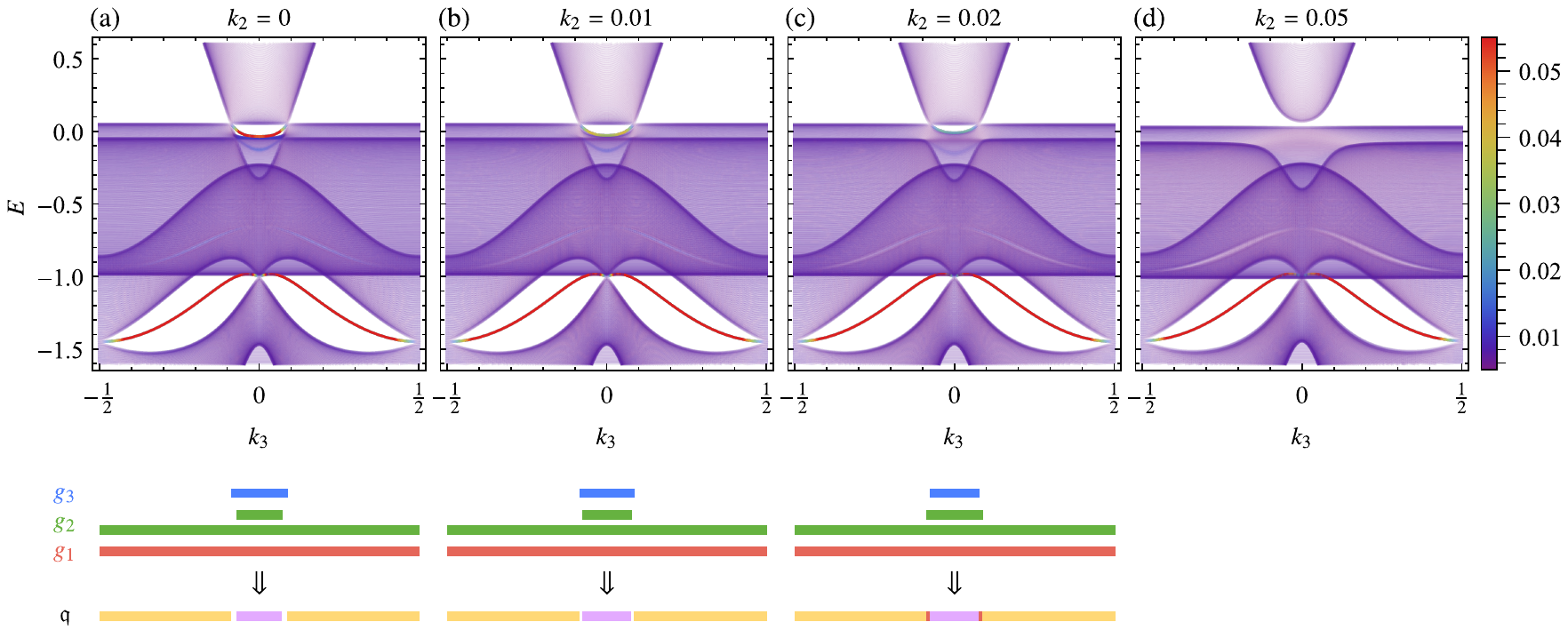}
    \caption{
    Band structure of the slab Hamiltonian of \ce{Li2NaN} defined in \cref{SM:eq:H_slab} for $N=200$ layers and $s=-0.05\nunit{eV}$, plotted for different values of $k_2$ corresponding to the four cuts through the SBZ shown in the right panel of \cref{SM:fig:li2nan:SBZQcharge}.
    The color scale indicates the inverse participation ratio (IPR), \cref{SM:eq:IPR}, according to the legend on the right.
    Below panels (a--c) the extent of each surface band is indicated by a colored horizontal bar, where the color indicates the gap in which the surface band occurs.
    The color scheme matches the one of \cref{SM:fig:li2nan:SBZQcharge}, which is indicated by the corresponding generators $g_j$.
    Multiplying the generators $g_j$ that are present at a given $k_3$ results in the quaternion charge, as shown in the bottom-most bar.
    The colors of this latter bar match \cref{SM:fig:li2nan:SBZQcharge} along the corresponding cut.
    We omit this discussion in panel (d) due to the difficulty to recognize the hybridized surface bands at this combination of $s$ and $k_2$.
    See the text of Sec.~\ref{SM:Subsec:bulk_boundary:surface_spectrum} for more details.
    }
    \label{SM:fig:li2nan:slab:Ek3k2}
\end{figure*}

We now discuss the correspondence between the bulk quaternion charges and the surface states.
The naively expected bulk-boundary correspondence~\cite{Wu:2019} suggests that surface states appear between bands $j$ and $j+1$ if the quaternion charge contains the generator $g_j$ as a factor (which corresponds to an inversion of those two bands).
Similar to \cref{fig:li2nan}(g) in the main text, we show the bulk quaternion charge for the closed paths projecting onto points in the SBZ in the absence of strain [\cref{SM:fig:li2nan:slab:Ek3k2_noStrain}(a)] and with a large tensile strain [\cref{SM:fig:li2nan:SBZQcharge}].

To start our comparison, we first consider the unstrained case, when the quaternion charge near $k_2 = k_3 = 0$ is $\pm g_1 g_2$~[\cref{SM:fig:li2nan:slab:Ek3k2_noStrain}(a)].
Note that the quaternion charge is not well-defined \emph{at} the line $k_2=0$, because red NLs project onto the SBZ there.
In accordance with the anticipated bulk-boundary correspondence, we find one surface band in both the first and the second energy gap [cf.\ \cref{SM:fig:li2nan:slab:Ek3k2_noStrain}(b,d)].
For another comparison, we consider a large strain value $s=-0.05\,\textrm{eV}$.
The quaternion charges over the whole SBZ are displayed in \cref{SM:fig:li2nan:SBZQcharge}, and we compare them to the surface states for various $k_2$ and $k_3$ [\cref{SM:fig:li2nan:slab:Ek3k2}].
We again find a correspondence with the anticipated bulk-boundary correspondence~\cite{Wu:2019}, however, one has to be careful with the counting and with the inclusion of the surface bands that hybridize with the bulk states, therefore we expound a few cases in more detail.

Referring to \cref{SM:fig:li2nan:slab:Ek3k2} for surface bands and to the right panel of \cref{SM:fig:li2nan:SBZQcharge} for the quaternion charge, we make the following observations.
In panels (a--c) there are four surface states states: one between bands $1\-- 2$, two of different extent along $k_3$ between bands $2\-- 3$ and one between $3\-- 4$.
First, for all four cuts the quaternion charge contains a factor $g_1$, which is consistent with the surface band in the gap $1\-- 2$ extending over all $k_3$ values.
Furthermore, the surface band in the gap $3\-- 4$ corresponds to a factor $g_3$ which appears inside the blue nodal ring in \cref{SM:fig:li2nan:SBZQcharge}, and thus the corresponding range of $k_3$ shrinks for increasing $k_2$ [panels (a--c)] until it vanishes completely [panel (d)].
Finally, for the surface bands in the $2\-- 3$ gap, the discussion becomes a bit more involved due to the missing bulk gap, which leads to hybridization of the possible surface states with the bulk states.
Nevertheless, we are able to identify two surface bands between these two bands, which are barely visible in panels (a--c), but not visible anymore in panel (d).
For values of $k_3$ where both of these surface bands are present, we expect a trivialization ($\pm g_2^2 \sim \pm 1$) and thus correspondence to \emph{no} factor $g_2$ in the quaternion charge, while $g_2$ should be present if there is only one of these two surface bands.
Note that in panels (a,b) the $2\-- 3$ surface band higher in energy has an extent in $k_3$ which is slightly smaller than the one of the $3\-- 4$ surface band, such that a (small) region with surface bands in all three gaps results.
This is consistent with the small white overlap shown in the inset of the right panel of \cref{SM:fig:li2nan:SBZQcharge}.
The white region has vanished in the third cut [panel (c)], which is reflected in the upper $2\-- 3$ surface band now having a larger extent in $k_3$ than the $3\-- 4$ surface band.

\section{Density of states and optical conductivity of \texorpdfstring{\NoCaseChange{\ce{Li2NaN}}}{Li2NaN}}\label{SM:Sec:DOS_cond}

In this section we elaborate on the calculations and results for the density of states and the optical conductivity of \ce{Li2NaN}, which have only been briefly discussed in the main text.
In preparation for these calculations, we compute the cumulative density of states for the TB model for various values of the strain parameter using the \Python{} package \WB{}~\cite{Tsirkin:2020, Liu:2020}.
We use an initial $\vec{k}$-space grid of $400^3$ points and perform $40$ iterations of adaptive refinement, cf.~\cite{SuppData} for the full \Python{} script and the explicity results.
By imposing charge neutrality, we can deduce the Fermi level $E_F$ for each case.

\subsection{Density of states}\label{SM:Subsec:li2nan:DOS}

\begin{figure}[b!]
    \centering
    \includegraphics{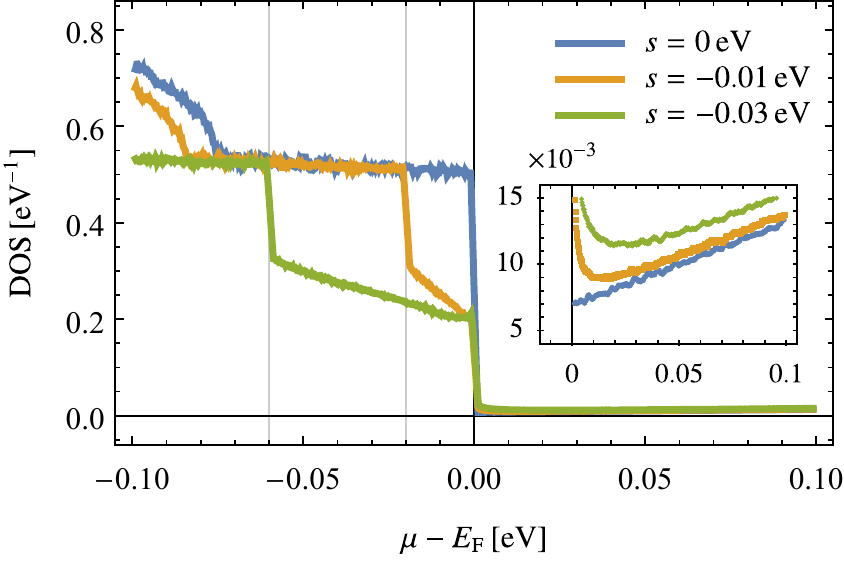}
    \caption{
        Density of states (DOS) of \ce{Li2NaN} for chemical potential $\mu$ near the Fermi level $E_\mathrm{F}$ obtained from the tight-binding model without and with strain.
        The inset shows a close-up of the jump at $\mu=E_\mathrm{F}$.
        The data is smoothed by a moving average over ten points. The same data, but plotted on a logarithmic scale, are included as \cref{fig:li2nan}(j) of the main text.
    }
    \label{SM:fig:li2nan:DOS}
\end{figure}

We verify the observation of a large jump in the density of states (DOS) near the Fermi level by computing it for the TB model using the \Python{} package \WB{}~\cite{Tsirkin:2020, Liu:2020}.
Starting with an initial $\vec{k}$-space grid of $400^3$ points we perform $2600$ iterations of adaptive refinement [cf.~\cite{SuppData} for the full \Python{} script].
The results of this calculation for $s\in\{0,-0.01,-0.03\}\nunit{eV}$ are presented in \cref{SM:fig:li2nan:DOS}.
Note that in reality the jumps would be smeared over a region of approximately $40\nunit{meV}$ due to the dispersion along $\Gamma A$ which is assumed to be completely flat in the TB model.

The single large jump from approximately $0.5\nunit{eV^{-1}}$ to $7\times 10^{-3}\nunit{eV}$ at the Fermi level for the unstrained case is split into two consecutive jumps at approximately $-2s$ and slightly above the Fermi level for the strained case.
Just as the single jump is explained by the flat dispersion of the two degenerate bands of the 2D IR along $\Gamma A$, the double jump follows from the same two bands after being split by the symmetry-breaking strain.
This is consistent with the distance between the two jumps being $2s$ and thus equal to the energy splitting of the nitrogen $p_{x,y}$ orbitals.

\subsection{Optical conductivity}\label{SM:Subsec:li2nan:cond}

For the calculation of the optical (interband) conductivity of \ce{Li2NaN} based on our TB model using the \Python{} package \WB{}~\cite{Tsirkin:2020, Liu:2020} we chose an initial grid of $625^3$ points in $\vec{k}$-space and ten iterations of adaptive refinement.
Furthermore, the Kubo formula requires us to choose a smearing parameter for approximating the delta function~\cite{Liu:2020} and we set that to $\eta=0.01\nunit{eV}$.
The full \Python{} script used to obtain the data is given in Ref.~\onlinecite{SuppData}.

Here we present additional results for the optical conductivity that are not shown in the main text.
We show both the real and the imaginary part of all non-vanishing components of $\sigma_{\alpha\beta}$.
Note that in the absence of strain $\sigma_{xx}=\sigma_{yy}$ due to the symmetries.
Similar to $\Re\sigma_{xx}$, which we discussed in the main text, $\Re\sigma_{yy}$ is strongly suppressed for small $\hbar\omega$ when increasing the strain parameter $s$.
However, while the peak of $\Re\sigma_{xx}$ near $\hbar\omega\approx 0.32\nunit{eV}$ is decreased when applying strain, it becomes more pronounced (and is shifted towards larger frequencies) for $\Re\sigma_{yy}$.
While the imaginary part can be obtained from the real part using the Kramers-Kronig relation, it highlights the differences of $\sigma_{xx}$ and $\sigma_{yy}$ under strain.
In both cases there is a large drop and thus a local minimum for finite but small frequencies, but in $\Im\sigma_{yy}$ this is accompanied by the formation of a sharp peak at higher frequencies.
This peak is located roughly at $\hbar\omega\approx 2\abs{s}$ and becomes very pronounced for large $s$.

\begin{figure*}[t!]
    \centering
    \includegraphics{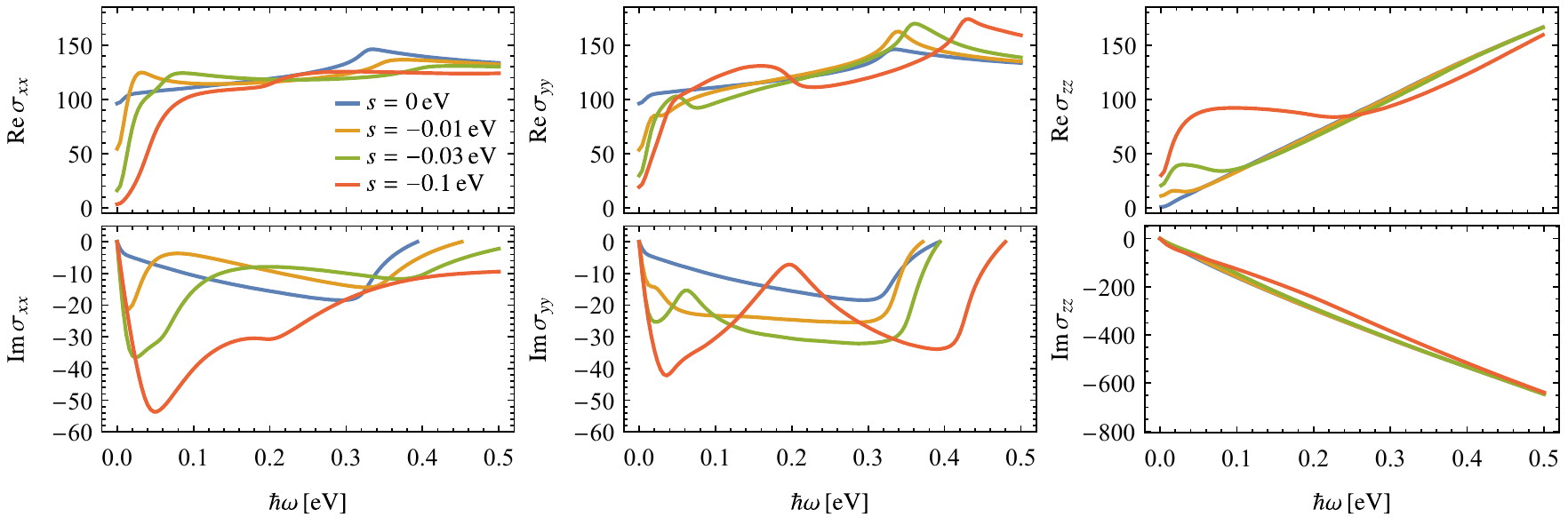}
    \caption{
        The real and the imaginary part of the interband optical conductivity in $\mathrm{S/cm}$ for different values of the strain parameter $s$ (cf.\ inset legend on the top left).
    }
    \label{SM:fig:li2nan:opt_cond}
\end{figure*}

\section{Triple-point materials}\label{SM:Sec:TP_materials}

\begin{figure*}
    \centering
    \includegraphics{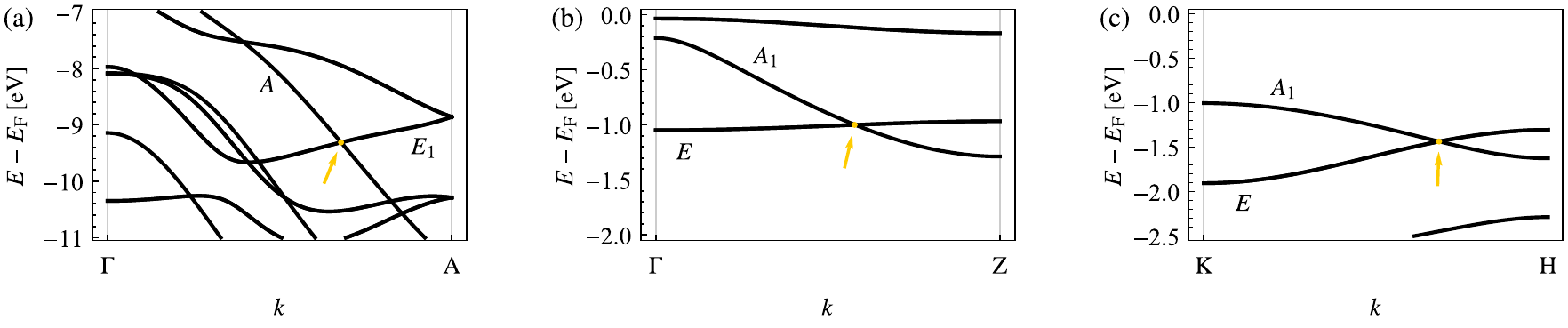}
    \caption{
        Band structure near the triple point (TP) and irreducible corepresentations of the corresponding bands along the relevant rotation-invariant line for (a) \ce{C3N4} (space group 176), (b) \ce{Na2LiN} (space group 129), and (c) \ce{MgH2O2} (space group 164).
        The relevant TP is marked by a yellow dot and arrow.
        Note that in some of the materials there are additional TPs present, some of them may be of a different type than the one marked. A detailed study of these materials appears in Ref.~\onlinecite{Lenggenhager:2022:TPClassif}.
    }
    \label{SM:fig:TPmaterials}
\end{figure*}

In the main text we present three materials exhibiting TPs that have not been reported previously [cf.\ compounds in \cref{tab:triplepoint_classif} marked with $\dagger$].
We identified these materials based on the symmetry criteria given by our classification and first-principles calculations [cf.\ \cref{SM:Subsec:tb_model:dft}].
\Cref{SM:fig:TPmaterials} shows the band structure near one of the TPs along the relevant rotation-invariant line in each of those compounds as well as the irreducible corepresentations of the involved bands.
This justifies the placement of those material candidates in \cref{tab:triplepoint_classif} of the main text.
We study the TPs of these materials and their associated attached nodal-line arcs in detail in a follow-up paper~\cite{Lenggenhager:2022:TPClassif}, where we explicitly verify the agreement with the presented classification of the TPs.

\end{bibunit}

\end{document}